\documentclass[reqno]{amsart}
\usepackage{amsmath}
\usepackage{amsthm}
\usepackage{amssymb}
\usepackage[
left=1.25in, right=1.25in]{geometry}
\usepackage{tikz}
\usepackage[all]{xy}
\usepackage{mathrsfs}
\usepackage{hyperref}
\usepackage{microtype}
\usepackage{url,doi}
\usepackage{braket}
\usepackage{breakurl}
\usepackage{ifpdf}
\usepackage{mathtools}
\usepackage{color,soul}

\usepackage[figuresright]{rotating}
\usepackage{bm}
\include{epsf}
\usepackage{color}
\usepackage{xcolor}
\usepackage{braket}
\usepackage{amsmath,amssymb,amsthm,mathrsfs,amsfonts,dsfont,bm,mathtools}  
\usepackage{graphicx}
\usepackage{subfigure}
\usepackage[titletoc]{appendix}
\usetikzlibrary{arrows,intersections}
\usepackage{verbatim}
\DeclareSymbolFont{symbolsC}{U}{pxsyc}{m}{n}
\DeclareMathSymbol{\coloneqq}{\mathrel}{symbolsC}{"42}

\newcommand{\C}{{\mathscr C}}
\newcommand{\Z}{{\mathscr Z}}
\newcommand{\F}{{\mathscr F}}

\newcounter{mycite}
\newtoks\citetoks
\makeatletter
\DeclareRobustCommand\unscite[1]{%
  \@ifundefined{uns@cite#1}
    {\refstepcounter{mycite}\label{citelabel@#1}%
     \expandafter\xdef\csname uns@cite#1\endcsname{\arabic{mycite}}%
     \toks\z@=\expandafter{\the\citetoks}%
     \toks\tw@=\expandafter\expandafter\expandafter{%
       \csname uns@bibitem#1\endcsname}%
     \edef\@tempcite{\the\toks\z@\the\toks\tw@}%
     \global\citetoks=\expandafter{\@tempcite}%
    }{}[\@nameuse{uns@cite#1}]}
\newcommand{\mybibitem}[2]{%
  \@namedef{uns@bibitem#1}{\bibitem[\ref{citelabel@#1}]{#1}#2}}
\makeatother
\usepackage{verbatim}
\usepackage{graphicx,adjustbox}

\begin{document}

\title{Exact Solvability and Asymptotic Aspects of generalized XX0 spin chains}

\author{M. Saeedian\textsuperscript{1}}
\address{\textsuperscript{1}School of Physics, Institute for Research in Fundamental Sciences (IPM), Tehran,
19395-5531, Iran}

\author{A. Zahabi\textsuperscript{2}}
\address{\textsuperscript{2}National Institute for Theoretical Physics,
School of Physics and Mandelstam Institute for Theoretical Physics,
University of the Witwatersrand,
Johannesburg Wits 2050,
South Africa}
\address{\textsuperscript{2}Centre for Research in String Theory, School of Physics and Astronomy, Queen Mary University of London, Mile End Road, London E1 4NS, UK}


\begin{abstract}
Building on our earlier work \unscite{Sa-Za}, we introduce and study generalized XX0 models. We explicitly construct a long-range interacting spin chain, referred to as the Selberg model, and study the correlation functions of the Selberg and XX0 models. Using a matrix integral representation of the generalized XX0 model and applying asymptotic analysis in non-intersecting Brownian motion, the phase structure of the Selberg model is determined. We find that tails of the Tracy-Widom distribution, of Gaussian unitary ensemble, govern a discrete-to-continuous third-order phase transition in Selberg model. The same method also reproduces the Gross-Witten phase transition of the original XX0 model. Finally, we conjecture universal features for the phase structure of the generalized XX0 model. \end{abstract}

\maketitle

\section{Introduction}
\label{Intro}
The XX0 spin chain, defined by the following Hamiltonian, as one of the simplest integrable and exactly solvable one-dimensional models, in the class of half-spin XYZ Heisenberg model on d-dimensional lattice, has been studied extensively
\unscite{Sa-Za},
\unscite{Schollwock},  \unscite{Bogoliubov}, \unscite{Its1}, \unscite{Its2}, \unscite{Its3}, \unscite{Its4}, \unscite{Izergin},
\begin{eqnarray}
\label{H_XX0 eq1}
{\hat{H}}_{XX0}=-\sum_{n=1}^{N}\sum_{m=1}^N{\Delta_{nm}}{\sigma }_{n}^{+}{\sigma }_{m}^{-},
\end{eqnarray}
where raising and lowering spin operators are defined as ${\sigma}_k^{\pm}=({\sigma}_k^{x}{\pm}i{\sigma}_k^{y})/2$ where ${\sigma}^{x,y,z}$ are the Pauli matrices and ${\Delta_{nm}}$ is the nearest neighbour coupling constant, satisfying the periodic boundary condition,
\begin{eqnarray}
\label{Delta eq2}
{\Delta_{nm}}=\frac{\Delta}{2}({\delta_{|n-m|,1}}+{\delta_{|n-m|, N-1}}).
\end{eqnarray}
Among the variety of classical, quantum and topological phase transitions that occur in XYZ models, the XX0 model does not experience those types of transitions. In contrast, because of a correspondence between the XX0 model and Non-Intersecting Brownian Motion (NIBM) \unscite{Bogoliubov}, this model faces a rather rare third-order phase transition \unscite{Sa-Za}, \unscite{Sc-Ma}. Similar phase transition in NIBM which is associated with the Douglas-Kazakov phase transition \unscite{Do-Ka} is studied in \unscite{Fo-Ma}. In our previous study, we explored the phase structure of XX0 spin chain and despite its simplicity, we found a sophisticated phase structure for this model \unscite{Sa-Za}. In this study, we generalize the model, in the simplest way, to a long-range interacting model and in the light of the simplicity of the model, we discuss a conjectured universality features in the phase structure of the generalized model arising from the conjectural universality of Tracy-Widom distribution (TW) \unscite{Tracy2}, \unscite{Tracy}, \unscite{Tracy3}.

In this work, we generalize the XX0 model into the most general extended XX0 model with Hamiltonian containing all possible short / long range interactions in the XX0 form, see Eq. \ref{General H}. This can be considered as the simplest long-range interacting spin chain. Long-range interacting spin chains are interesting generalizations of the known short-range models and they are studied to some extent, for example see \unscite{Bernard}, \unscite{Lipkin1}, \unscite{Lipkin2}, \unscite{Lipkin3}. The XX0 model as the simplest model of spin chains provides probably the simplest framework for studying the generalization of spin chains to long-range interacting models. Thus, generalizing the obtained phase structure of XX0 model to the generalized long-range model would be of great interest. Among the important motivations for this study, we can mention the test of the robustness of the phase structure of the XX0 model under the generalization. Among other motivations, one can look for exact solvability and integrable probability, meaning that there exist closed exact formulas for the partition functions and correlation functions for the generalized model. In this way, we start from an explicit example, we consider a specific long-range model called the Selberg spin chain model that turns out to be exactly solvable. Using these results we elaborate on the exact solvability of the generalized XX0 model and we obtain expansion formulas for the partition functions and correlation functions of the generalized model as well as its examples. In order to study the generalized XX0 model, we extend the methods from random matrix theory and integrable probability \unscite{Bogoliubov}, \unscite{Baik}, which are introduced and adopted to the infinite XX0 model in our previous study \unscite{Sa-Za}. We will collect the necessary background and results from these studies in a separate section at the end of the introduction.

Similar to the original model, the observables i.e. the partition function and correlation functions, of the generalized model have matrix integral representation, and the matrix integral is analytically solvable and can be expanded and expressed in terms of Selberg integrals and thus the correlation functions have closed formulas in terms of Gamma functions. Moreover, the observables of the generalized models
have the determinantal structure \unscite{Borodin}, i.e. the Toeplitz Determinant (TD) and Fredholm Determinant (FD), which plays the crucial role in the integrability of the model as well as in the asymptotic analysis and exploring the phase structure of the model. However, to prevent prolonging the article, we have collected all these results in the appendices (C) and (D). 

Selberg integral \unscite{Forrester1} and FD \unscite{Borodin} representations provide new exact closed expressions for the observables of the finite- and infinite-size generalized model. As our first new result, by using i) expansion of the Selberg potential as power series, ii) specification of the Selberg potential to write other potentials and iii) Schur function factorization property of the Selberg integral,  we show that some of the partition functions and correlation functions of the infinitely large generalized model and its specific examples, including the original XX0 model, Selberg models, quadratic model, single-term Hamiltonian model, etc. can be expressed in terms of the Selberg integrals and thus Gamma functions. Then, using the results indicating the relation between TD and FD \unscite{Bo-Ok}, we show that the partition function of both finite- and infinite-size general models have closed determinantal formulas, i.e. FD.  

Beyond these formal expressions and the exact solvability of the model, FD formula via the Riemann-Hilbert problem makes the asymptotic analysis of the generalized model a feasible problem and in fact, as we explicitly see in the case of the Selberg model, the limit of the FD with Airy kernel leads to TW. In the physics language, the asymptotic analysis is the study of the model in the thermodynamic limit, in which the phase structure of the model emerges and thus we expect that the TW determines all the information relevant to the phase structure. Applying the introduced method in \unscite{Sa-Za}, we compute the free energy of the finite model in the asymptotic limit from the TW and thus extract the phase structure of the model. In the case of the original XX0 model, we apply this new spirit and method from TW to reproduce the known results of the Gross-Witten (GW) phase transition \unscite{Gross}, namely the free energy near the phase transition point. As an example of a long-range interacting model, we apply the same technique to the Selberg model and compute the free energy and find a new phase structure for this model. Similar to the original model, we obtain the third-order phase transition between finite and infinite size model with a similar domain wall of the same order. For the sake of completeness, we review the obtained results in our previous study for the original model and its weak coupling limit, and compare them with the Selberg model.

The aforementioned dynamics ruled by Eq. (1) is special case of the asymmetric exclusion process (ASEP) (see Section 2.3 in \unscite{Hinrichsen}). In this many-body particle system, the hard-core particles walk randomly into the unoccupied neighbor sites with the rate $q$ $(q-1)$ to the left (right) direction in the one-dimensional lattice. The exclusion principle puts a constraint on the dynamics such that at any time in each site of the lattice there is at most one particle. In special case $q=1$, the ASEP becomes non-intersecting Brownian motion and thus its dynamics is that of the XX0 model. As we observed in \cite{Sa-Za}, the asymptotic behavior of the free energy, i.e. the order parameter in this model, indicates that the XX0 model undergoes a third-order phase transition, called Gross-Witten phase transition. Adopting this result for the non-equilibrium critical phenomena in $(q=1)$ ASEP, suggests exact new solutions for the stationary states of the dynamics of the ASEP at zero temperature.


\section*{Essentials of matrix model/xx0 model}
The essential observations in the studies \unscite{Bogoliubov}, \unscite{Sa-Za} can be summarized in the following formulas for partition functions and correlation functions. In the limit of large size, $N\gg 1$, the partition function is given by
\begin{eqnarray}
\label{matrix continous}
\Z_{XX0}&\coloneqq&\bra{\Uparrow}\prod_{i=1}^{N_f}\sigma^{+}_{N_f-i}e^{-t{\hat{H}}_{XX0}}\prod_{i=1}^{N_f}\sigma^{-}_{N_f-i}\ket{\Uparrow}\coloneqq\bra{\overbrace{{\uparrow},...,{\uparrow},\underbrace{{\downarrow},...,{\downarrow},{\downarrow}}_{N_f}}^{N}}e^{-t{\hat{H}}_{XX0}}\ket{\overbrace{\underbrace{{\downarrow},{\downarrow},...,{\downarrow}}_{N_f},{\uparrow},...,{\uparrow}}^{N}}\nonumber\\
&=&\prod_{j=1}^{N_{f}} \int_{-\pi}^{\pi} \frac{d\alpha_{j}}{2\pi} f_{GW}(e^{i\alpha_{j}}) \prod_{l<p} | e^{\imath\alpha_{l}}-e^{\imath\alpha_{p}}|^{2}\nonumber\\
&=&\det\begin{bmatrix}\int_{|z|=1}z^{-j+l}f_{GW}(z)\frac{dz}{2\pi iz}\end{bmatrix}_{j,l=0}^{N_{f}-1}\nonumber\\
&\coloneqq&D_{N_{f}}(f_{GW}),
\end{eqnarray}
where $\ket{\Uparrow}=\otimes_ {n=1}^{N}\ket{\uparrow}_{n}$ denotes the ferromagnetic vacuum, \textit{t} is the evolutionary parameter or time and $N_f$ is the number of magnons in the ferromagnetic vacuum, $f_{GW}=e^{tV_{GW}}$ is the weight function of the Gross-Witten potential $V_{GW}(z)=\frac{z+z^{-1}}{2}$ and $z=e^{\imath\alpha}$ ($\alpha$ is eigenvalue of random matrix).
The general correlation functions and their matrix integral representation are obtained in \unscite{Bogoliubov},
\begin{equation}
\label{C}
\C_{j_{1},...,j_{N_{f}};l_{1},...,l_{N_{f}}}\coloneqq\bra{\Uparrow}\sigma^{+}_{j_{1}}...\sigma^{+}_{j_{N_{f}}}e^{-tH_{XX0}}\sigma^{-}_{l_{1}}...\sigma^{-}_{l_{N_{f}}}\ket{\Uparrow}=
\end{equation}

\begin{eqnarray}
\label{CM}
\frac{1}{N_f!}\bigg( \frac{1}{2\pi} \bigg)^{N_{f}}\int^{\pi}_{-\pi}d\alpha_{1}...
                                            \int^{\pi}_{-\pi}d\alpha_{N_{f}}&&
                               S_{\lambda}(e^{-\imath\alpha_{1}},e^{-\imath\alpha_{2}},...,e^{-\imath\alpha_{N_{f}}})
                                S_{\lambda'}(e^{\imath\alpha_{1}},e^{\imath\alpha_{2}},...,e^{\imath\alpha_{N_{f}}})\nonumber\\
                                            &&e^{t\sum_{m=1}^{N_{f}}cos_{}\alpha_{m}}\prod_{1\leq j < k \leq N_{f}}|e^{\imath\alpha_{j}}-e^{\imath\alpha_{k}}|^{2},
\end{eqnarray}
where $\sigma^{\pm}$ are the spin operators act on $l_{1},...,l_{N_{f}}$ ($j_{1},...,j_{N_{f}}$) positions in the initial (final) ferromagnetic state and flip $N_f$ spins in these positions, $S_\lambda$ is the symmetric Schur function of strict partition $\lambda=(N\geq\lambda_1>\lambda_2>...>\lambda_{N_f}\geq 0)$ with $\lambda_i=j_i-N_f+i$ and $\lambda'_i=l_i-N_f+i$ ($i=0, 1, ..., N_f-1$). 

The rest of the paper is organized as follows. In chapter two, the generalized XX0 model introduced and its matrix integral and TD representations are presented. In chapter three, we explore the phase structure of the XX0 model and a new generalized XX0 long-range interacting spin chain, called the Selberg model, by using TW. In the discussion, we formulate our conjecture about the universality of the phase structure in the generalized model. There are appendices for the mathematical details of some of the new results of this paper and reviewing the known mathematical results that are useful in this study. Especially, in Appendices (C) and (D), we compute the partition function and correlation functions of the generalized model and its examples in terms of the Selberg integrals and FD.
\section{Generalized XX0 model and matrix integral representation}
\label{G XX0}
In this part, we define generalized XX0 spin chain with interaction terms between spins separated by an arbitrary distance. This generalization provides mathematical and physical frameworks to examine the ideas and techniques which are successfully used for the original XX0 model. Then, the partition function and the correlation functions in the generalized XX0 model are represented in terms of matrix integral and TD. These representations allow us to use the results from NIBM \unscite{Baik}.

We start with the periodic finite size one-dimensional quantum spin model with generalized Hamiltonian of the XX0 type, which includes arbitrary length interactions \cite{Tierz},
\begin{equation}
\label{General H} 
\hat{H}_{Gen}=-\sum_{n=0}^{N}\sum_{m=1}^{N/2}\Delta_{m}\Big(\sigma_{n}^{+}\sigma_{n+m}^{-}+\sigma_{n}^{+}\sigma_{n-m}^{-}\Big),
\end{equation}
where for simplicity, \textit{N+1} which is the size of chain is taken to  be odd, \textit{n} and \textit{m} indicate the position of spins in the one-dimensional lattice and we choose the middle site of the lattice to be indexed by $n=0$, $\Delta_m$ is the interaction coupling between spins at $n$ and $m$. This spin chain is defined and studied for the first time in \unscite{Tierz}. The case of $\Delta_ m=\delta_{1,m}$ corresponds to the familiar XX0 Hamiltonian \unscite{Lieb}, \unscite{Sa-Za}. The Hamiltonian (\ref{General H}) can recognize a magnon located at any position in the ferromagnetic vacuum via interaction terms with all possible lengths in the chain.

In this work, we study the generalized XX0 spin with the generalized Hamiltonian and its explicit examples with zero-, short-, long-, infinite-range interactions. We continue with the observable of the model.
The arbitrary time correlation function of the model including an arbitrary number of magnons in arbitrary positions in the vacuum and the partition function of the model which is defined as a specific arrangement of magnons are written as,
\begin{eqnarray}
\label{Z_sandwich}
\Z_{Gen}^{d}\coloneqq\bra{\Uparrow}\prod_{j=1}^{N_f}\sigma^{+}_{N_f-j}e^{-t{\hat{H}}_{Gen}}\prod_{i=1}^{N_f}\sigma^{-}_{N_f-i}\ket{\Uparrow},\quad \C_{Gen}^{d}\coloneqq\bra{\Uparrow}\prod_{i=1}^{N_f}\sigma^{+}_{j_i}e^{-t{\hat{H}}_{Gen}}\prod_{i=1}^{N_f}\sigma^{-}_{l_i}\ket{\Uparrow}.
\end{eqnarray}
In fact, the defined partition function plays a similar role to the order parameter in our study and determines the phase transition in the generalized XX0 model. For more explanations and justifications for the definition of the partition function see \unscite{Sa-Za}.

By using the operator formalism and the Bessel functions in the XX0 model \unscite{Bogoliubov}, and in the generalized XX0 model \unscite{Tierz}, \unscite{Sa-Za}, it has been shown that the correlation functions of the finite-size generalized XX0 model can be written in the discrete matrix integral representation and also by using the Heine-Szeg\"{o} identity as discrete TD on domain $d= \{z \in \mathds{C}  : z^{m} = 1\}$ with size $|d|=m$ as, 
\begin{eqnarray}
\label{matrix discrete}
\C_{Gen}^{d}&=&c\sum_{N>s_1>s_2>...>s_{N_f}\geq 0}\prod_{j=1}^{N_{f}}   f_{Gen}(e^{\imath\alpha_{s_j}})
S_{\mu}(e^{-\imath\alpha_1},...,e^{-\imath\alpha_{N_{f}}})
S_{\lambda}(e^{\imath\alpha_1},...,e^{\imath\alpha_{N_{f}}})
\prod_{l<p} | e^{\imath\alpha_{l}}-e^{\imath\alpha_{p}}|^{2}\nonumber\\
&=& \det\begin{bmatrix}\frac{1}{|d|} \sum_{z \in d} z^{-j+l}f_{Gen}(z)S_{\mu}(z^{-1})
S_{\lambda}(z)\end{bmatrix}_{j,l=0}^{N_{f}-1},
\end{eqnarray}
where $c=\frac{1}{(N+1)^{N_f}}$, $S_{\lambda}(e^{\imath\alpha})$ is the Schur function indexed by partition $\lambda$, with entries $\alpha_i$ eigenvalues of orthogonal random matrix ensemble \unscite{Mehta org}, $\lambda_i=l_i-N_f+i$, $\mu_i=j_i-N_f+i$, and the weight function is given by  

\begin{equation}
\label{V}
f_{Gen}(z)=e^{tV_{Gen}(z)}\quad,\quad V_{Gen}(z)=\sum_{m=1}^{\frac{N}{2}}\Delta_m\Big(z^{m}+z^{-m}\Big), 
\end{equation}
where $z=e^{\imath \alpha}$. Once we choose the specific arrangement of magnons as in l.h.s of Eq.~\ref{Z_sandwich}, by putting $S_\lambda=S_\mu=1$ for $\lambda=\mu=0$ in the matrix model representation of the correlation function in Eq.~\ref{matrix discrete}, the correlation function reduces to the partition function $\Z^d_{Gen}$. Furthermore, as we will see in the next section, the Schur functions factor out of the Selberg matrix integrals Eqs. \ref{Jack0} and \ref{Jack00}, and similarly in the Selberg model with potential \ref{Sel pot} and other models. Therefore, the correlation functions can be further simplified in terms of the Selberg integrals. Comparing Eq.~\ref{V} and Eq.~\ref{General H}, one can observe, \unscite{Tierz}, that the symmetric interactions with right and left $m$th neighbours in the spin chain correspond to the terms $\Delta_m z^m$ and $\Delta_m z^{-m}$ of the potential in the matrix integral representation.

In the limit of large parameters, for the infinite spin chain ($N\rightarrow\infty$), with appropriate scaling, one obtains the continuous matrix integral and TD for the partition function and correlation functions,
\begin{eqnarray}
\label{matrix continous}
\C_{Gen}&=&\prod_{j=1}^{N_{f}} \int_{-\pi}^{\pi} \frac{d\alpha_{j}}{2\pi} f_{Gen}(e^{\imath\alpha_{j}})
S_{\mu}(e^{-\imath\alpha_1},...,e^{-\imath\alpha_{N_{f}}})
S_{\lambda}(e^{\imath\alpha_1},...,e^{\imath\alpha_{N_{f}}})
\prod_{l<p} | e^{\imath\alpha_{l}}-e^{\imath\alpha_{p}}|^{2}\nonumber\\
&=& \det\begin{bmatrix}\int_{|z|=1}z^{-j+l}f_{Gen}(z)S_{\mu}(z^{-1})
S_{\lambda}(z)\frac{dz}{2\pi iz}\end{bmatrix}_{j,l=0}^{N_{f}-1}.
\end{eqnarray}
In this paper, we study the generalized XX0 model with Hamiltonian \ref{General H} and the corresponding matrix model with potential ~(\ref{V}). Throughout the article, we provide explicit calculations and results for the generalized model as well as following special cases of the generalized model:
\begin{itemize}
 \item Non-interacting XX0/matrix model with zero (constant) potential; $\Delta_{m} =0$.
 
 \item Nearest-neighbour interacting (original) XX0 model / matrix model with Gross-Witten potential, $V=z+z^{-1}$; $\Delta_{m}=c\delta_{m,1}$ and $c=1$.
 
 \item Weakly coupled XX0 model / Gaussian matrix model with quadratic potential $V_{Quadratic}=-z^2$; $\Delta_{m}=c\delta_{m,1}$ and $c\ll 1 $, \unscite{Sa-Za}.
 
 \item Single-term arbitrary-range interacting XX0 model/ matrix model with potential $V=\Delta_m (z^m+z^{-m})$.
 
 \item Specific long(infinite)-range XX0 model/Selberg matrix model as defined below.
 \end{itemize} 
\subsection{Selberg-XX0 model and observables}
The Selberg matrix model is characterized by $f_{Selb}(z)=e^{tV_{Selb}(z)}$ and the potential in L.H.S. Eq. \ref{Sel pot}. The Selberg potential can be expanded, as it is obtained in the Appendix.~\ref{APP G P}, and it can be written in the form of the potential in (R.H.S.) Eq.~\ref{V} in the limit $N\rightarrow \infty$ and with the coupling constant in R.H.S. Eq. \ref{Sel pot},  
\begin{equation}
\label{Sel pot}
V_{Selb}(z)=\frac{1}{t}\log\big(z^{-t}(1+z)^{2t}\big),\quad \Delta^{(Selb)}_m= \sum_{i=0}^{\big[\frac{N-m}{2}\big]}\frac{(-1)^{(2i+m)-1}}{2^{(2i+m)}(2i+m)}\binom{2i+m}{i}.
\end{equation}
The Selberg XX0 model is defined by the Hamiltonian \ref{General H} with the coupling given by R.H.S. Eq. \ref{Sel pot}. This is a specific generalization of the XX0 model with infinite/long-range interaction and with the specific values of the coupling which is tending to zero, $\Delta_m\rightarrow 0$ as $m\rightarrow \infty$. However, the finite-size Selberg model, or the Selberg model with finite-size effects, can only make sense in the asymptotic limit with infinitely large size since the range of interaction is infinite. The Selberg model should be thought of as a model with infinitely large size which has finite-size effects in definite regions of the moduli space, as this will be explained in section 3.

After defining the model, the next step is to study the observables of the model, i.e. the correlation functions.
Before we focus on the phase structure of the general XX0 model and the universality conjecture, by using the theory of Selberg and Mehta integrals we elaborate on the exact solvability of the model via obtaining closed formulas for correlation functions in different cases. 

By using the introduced necessary techniques, the Selberg integral and the Schur function factorization property, see Appendix C.1, we will compute the partition functions and correlation functions in each case. By tuning parameters $a$ and $b$, such that $a=b=t$, $t>0$, and $\gamma=1$ in the generalized Selberg integral \ref{Selber} in Appendix C, and using the matrix integral representation of the generalized XX0 model Eq.~\ref{matrix continous}, the partition function of the Selberg XX0 model, can be obtained as,
\begin{eqnarray}
\label{Selber2}
\Z_{Selb}&=&\bra{\Uparrow}\prod_{j=1}^{N_f}\sigma^{+}_{N_f-j}e^{-t{\hat{H}}_{Selb}}\prod_{i=1}^{N_f}\sigma^{-}_{N_f-i}\ket{\Uparrow}\nonumber\\
&=&\frac{1}{(2\pi)^{N_f}}\int_{-\pi}^{\pi}...\int_{-\pi}^{\pi}\prod_{i=1}^{N_f}e^{-\imath t\theta_i}(1+e^{\imath\theta_i})^{2t} \prod_{1\leq i< j\leq N_f}|e^{\imath\theta_i}-e^{\imath\theta_j}|^{2}d\theta_1...d\theta_{N_f}\nonumber \\
&=&\prod_{j=0}^{N_f-1}\frac{\Gamma(1+2t+j)\Gamma(2+j)}{\Gamma(1+t+j)^2\Gamma(2)}\nonumber \\
&=&D_{N_f}(f_{Selb}).
\end{eqnarray}
Similarly, for the finite-size model we have
\begin{equation}
\Z_{Selb}^d=\det\begin{bmatrix}\frac{1}{|d|} \sum_{z \in d} z^{-j+l}f_{Selb}(z)\end{bmatrix}_{j,l=0}^{N_{f}-1}=D^{d}_{N_f}(f_{Selb}, |d|). 
\end{equation}

In the next step, we compute the correlation functions of generalized XX0 models, using Eq.~\ref{matrix continous} and the identities \ref{Jack0} and \ref{Jack00} for the Schur functions in the Selbeg integral.
Similar to the partition functions, the correlation functions of the model with a given potential can be obtained from Eqs.~\ref{Jack0} and \ref{Jack00} by tuning its parameters $a$, $b$.
A specific case of the correlation function of the Selberg model with fixed left state, the same as the state in partition function, can be obtained from the Selberg integral with one Schur function,  Eq.~\ref{Jack0},
\begin{eqnarray}
\label{factoring} \nonumber
\C_{Selb}^*&=&\bra{\Uparrow}\prod_{i=1}^{N_f}\sigma^{+}_{N_f-j}e^{-t{\hat{H}}_{Selb}}\prod_{i=1}^{N_f}\sigma^{-}_{l_i}\ket{\Uparrow}\nonumber\\
&=&\frac{1}{(2\pi)^{N_f}}\prod_{i=1}^{N_f}\int_{-\pi}^{\pi}d\theta_iS_{\lambda}(e^{\imath\theta})e^{-\imath t\theta_i}(1+e^{\imath\theta_i})^{2t}\prod_{1\leq i< j\leq N_f}|e^{\imath\theta_i}-e^{\imath\theta_j}|^{2}\nonumber \\
&=&\frac{(-1)^{-|\lambda|}[-t]_{\lambda}}{[t+N_f]_{\lambda}}S_{\lambda}(1^{N_f})\prod_{j=0}^{N_f-1}\frac{\Gamma(1+2t+j)\Gamma(2+j)}{\Gamma(1+t+j)^2\Gamma(2)}\nonumber\\
&=&\frac{(-1)^{-|\lambda|}[-t]_{\lambda}}{[t+N_f]_{\lambda}}S_{\lambda}(1^{N_f})\Z_{Selb},
\end{eqnarray}
wherein the third line we used the fact that the Schur function $S_{\lambda}$ is a homogeneous polynomial of degree $|\lambda|$.
Thus, in the Selberg case, we observe that the Schur function completely factor out from the expression and the correlation function becomes proportional to the partition function.  Based on currently available techniques, the full correlation function, i.e. the Selberg integral with two Schur functions does not allow for an explicit evaluation. Similar to the above Selberg integral representation of the partition function and correlations, the partition function and correlations of the original XX0 model and other related models can be written as infinite sums over Selberg integrals. We have collected some of the results in Appendix (\ref{C_P}). In the Appendices (C) and (D),
explicit formulas for the partition functions and correlations of the finite- and infinite-size generalized model, and its specific examples are obtained in terms of the Selberg integrals and Fredholm determinants.

In the following sections, based on the matrix integral representation and bijection with NIBM, we study the phase structure of the generalized XX0 model in some specific cases, by using the FD representations of the partition functions which is a suitable representation to study the asymptotic limit. 

\section{Phase structure of the Selberg XX0 model; main results}
In this section, we obtain the free energies of the generalized XX0 models in some special cases, by using TW technique. Then, using explicit formulas for free energy, we explore the phase structure of the models and extract possible phase transitions.
We study the possible phase transitions in both finite and infinite chain, with long- or short-range interactions, by focusing on the TW distribution as a driving force and essential element behind these transitions.

As we mentioned, the importance of the FD representation is in its crucial role in the integrability and also the asymptotic analysis of spin chains.
In the next step, we adopt and use the FD formulas to study the asymptotic limit of the partition function and calculate the free energy and thus extract the phase structure of generalized XX0 spin chains. One of our goals in this work is to propose this method as an effective way to explore the phase structure of the model which besides the new results it generates, it reproduces the known results from the standard methods of random matrix models.


From the TW perspective, we can extract the phase structure of the model by using three observations and computations. First, the partition function of the finite/infinite model expressed as a FD \unscite{Bo-Ok},\unscite{Baik}. Second, the asymptotic result of the partition function (FD) as a TW distribution with an argument as a parameter, depending on the potential and/or the Hamiltonian of the model. The third step is to look at the asymptotic behavior of the TW distribution which is non-symmetric and it has different left and right tails, and calculate the free energy from the asymptotic behavior of the tails of the distribution. This procedure can be explicitly performed for the generalized XX0 model in some examples and the results uncover the phase structure of the model. In fact, the asymptotics of the TW is in charge of the corrections to the asymptotic limit of the partition function and since left and right tails of TW have different asymptotics, they lead to different corrections and thus a phase transition happens in the model.

An important observable of the model, the free energy is defined via the logarithm of the partition function in the appropriate asymptotic limit, ${\F}_{Gen}=\lim_{N_{f}\to\infty}\frac{1}{N_{f}^{2}}\log \Z_{Gen}$. The free energy of the infinite-size models can be obtained from the asymptotic limit ($N_f\rightarrow \infty$) of the results for partition functions in chapter two. However, the free energy of the finite-size model requires more involved asymptotic analysis as we will see later in this chapter. 

In the following, we obtain the free energy of the finite-size XX0-type spin chains in the asymptotic limit and our focus is to extract the phase structure of the model from the TW distribution.
\subsection*{Selberg XX0 model}
\label{G P} 
In this part, we consider a new example, a model with long(infinite)-range interaction and we follow the procedure introduced in \unscite{Za} to study its phase structure. Despite the infinite range of interaction, as we will show, this model undergoes a discrete-to-continuous phase transition.

Using the Selberg integral in the Appendix.~\ref{schur function}, the free energy of the infinite Selberg XX0 model is obtained in Eq.~\ref{Selber2}. Using the results obtained in \unscite{Baik}, putting together Eqs. \ref{con prob1} for $s=1$, and \ref{con prob2} in Appendix \ref{MIG}, we can obtain the partition function of the finite-size model from
\begin{eqnarray}
\label{Z/Z F Gw} 
\lim_{min(N, N_{f}, t)\to\infty}c_{Selb}\frac{\Z_{Selb}^{d}}{\Z_{Selb}}=F(x_{Selb}),
\end{eqnarray}
where $F(x)$ is the TW distribution function and $c_{Selb}=1/2\pi$ and 
$x_{Selb}=\frac{N-2\sqrt{N_{f}^{2}+2N_{f}t}}{(N_{f}^{2}+2N_{f}t)^{-\frac{1}{6}}t^{\frac{2}{3}}}$.

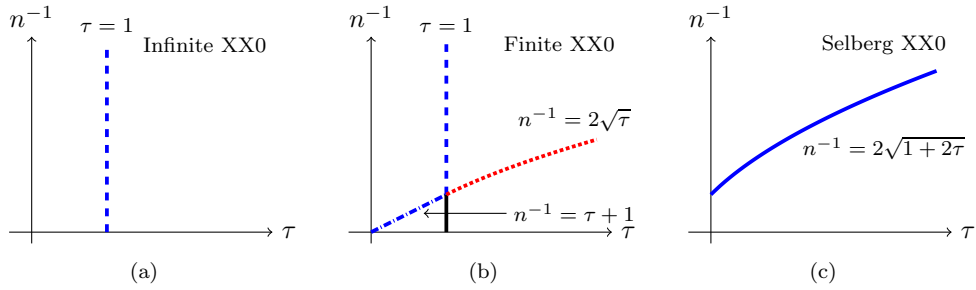
\begin{figure}[t!]
\centering

    \subfigure[]{
        \centering
        \pagestyle{empty}
         \begin{tikzpicture}[yscale=0.5,xscale=1]
          \draw[->] (-.3,1) -- (3.2,1) node[right] {$\tau$};
           \draw[->] (0,.7) -- (0,6.2) node[above] {$n^{-1}$};
           \draw[dashed][scale=1,domain=1:6,smooth,variable=\x,blue,line width=0.5mm] plot ({1},{\x});
           \node at (2.3,6) {\footnotesize Infinite XX0 };
           \node at (1,6.5) {\footnotesize $\tau=1$ };
         \end{tikzpicture}
        \label{f1}
    }
    ~
    \subfigure[]{
        \centering
        \pagestyle{empty}
         \begin{tikzpicture}[yscale=0.5,xscale=1]
            \draw[->] (-.3,1) -- (3.2,1) node[right] {$\tau$};
            \draw[->] (0,.7) -- (0,6.2) node[above] {$n^{-1}$};
            \draw[densely dashdotted][scale=1,domain=0:1,smooth,variable=\x,blue,line width=0.5mm] plot ({\x},{\x+1});
            \draw[dashed][scale=1,domain=2:6,smooth,variable=\x,blue,line width=0.5mm] plot ({1},{\x});
            \draw[][scale=1,domain=1:2,smooth,variable=\x,line width=0.5mm] plot ({1},{\x});
           \draw[scale=1,domain=1:3,smooth,variable=\x,red,densely dotted,line width=0.5mm] plot ({\x},{2*(\x)^(0.5)});
       
           \node at (2.5,6) {\footnotesize {Finite XX0} };
           \node at (1,6.5) {\footnotesize $\tau=1$ };
           \node at (2.7,4) {\footnotesize $n^{-1}=2\sqrt{\tau}$ };
           \node at (2.7,1.5) {\footnotesize $n^{-1}=\tau+1$};
           \draw[->] (1.8,1.5) -- (.7,1.5);
         \end{tikzpicture}
        \label{f2}
    }
    ~
    \subfigure[]{
        \centering
        \pagestyle{empty}
         \begin{tikzpicture}[yscale=0.5,xscale=1]
          \draw[->] (-.3,1) -- (3.2,1) node[right] {$\tau$};
           \draw[->] (0,.7) -- (0,6.2) node[above] {$n^{-1}$};
           \draw[scale=1,domain=0:3,smooth,variable=\x,blue,line width=0.5mm] plot ({\x},{2*(1+2*\x)^(0.5)});
       
           \node at (2.3,6) {\footnotesize Selberg XX0};
           \node at (2.3,3.25) {\footnotesize $n^{-1}=2\sqrt{1+2\tau}$};
         \end{tikzpicture}
        \label{f3}
    }
\caption{Phase structure of (a) infinite XX0 model, (b) finite-size XX0 model and (c) Selberg XX0 model.
}\label{F}
\end{figure}

The free energy of the infinite spin chain is defined as ${\F}_{Selb}=\lim_{N,N_{f},t\to \infty }\frac{1}{N_{f}^{2}}\log \Z_{Selb}$,
and by using the Selberg integral, Eq. \ref{Selber2},
\begin{eqnarray}
\label{G Free energy}
\Z_{Selb}=\prod_{j=1}^{N_{f}} \int_{-\infty}^{\infty} \frac{dz_{j}}{2\pi} z^{-t}(1+z)^{2t} \prod_{1\leq l< p\leq N_{f}} | z_{l}-z_{p}|^{2}=
\prod_{j=1}^{N_f}\frac{\Gamma(1+2t+j)\Gamma(2+j)}{\Gamma(1+t+j)^2\Gamma(2)},
\end{eqnarray}
and thus the free energy of the finite model can be obtained from the ratio formula \ref{Z/Z F Gw} as
\begin{eqnarray}
\label{G free energy0} 
\F_{Selb}^{d}=\F_{Selb}+\frac{1}{N_{f}^{2}}\log  F(x_{Selb})-\frac{1}{N_{f}^{2}}\log c_{Selb},
\end{eqnarray}
where ${\F}_{Selb}^{d}=\lim_{N,N_{f},t\to \infty }\frac{1}{N_{f}^{2}}\log \Z_{Selb}^{d}$. By fixing 
$\tau=\frac{t}{N_{f}}$ and $n^{-1}=\frac{N}{N_{f}}$ in the asymptotic limit, the argument of the TW becomes $x= j N_{f}^{2/3}$ 
with $j=\frac{n^{-1}-2\sqrt{1+2\tau}}{(1+2\tau)^{-\frac{1}{6}}\tau^{\frac{2}{3}}}$. 
Thus, the asymptotics of TW, \unscite{Ba-Di} adopted for the Selberg model, in terms of $n^{-1}$ and $\tau$ can be written
\begin{equation}
 \label{asym TW inf}
 F(n^{-1},\tau) = 
  \begin{cases}
   1-\mathcal{O}\big(e^{-j^{3/2}N_{f}}\big) & \quad n^{-1}>2\sqrt{1+2\tau} \\
   \mathcal{O}\big(e^{-|j|^3N_{f}^{2}}) & \quad n^{-1}<2\sqrt{1+2\tau} \\
  \end{cases}.
\end{equation}
Finally, inserting Eq.~\ref{G Free energy} and precise version of asymptotic results for TW \unscite{Ba-Di} in the case of Eq. \ref{asym TW inf}, in Eq.~\ref{G free energy0}, we can write the free energy of the Selberg spin chain as
\begin{eqnarray}
 \label{G free enrgy2 log}
 {\F}^{d}_{Selb}=  
  \begin{cases}
   {\F}_{Selb}+\frac{1}{N_{f}^{2}}\log (1-\frac{e^{-c_{1}j^{3/2}N_{f}}}{32\pi j^{3/2}N_{f}})-\frac{1}{N_{f}^{2}}\log(c_{Selb}) & \quad n^{-1}>2\sqrt{1+2\tau} \\
   \\
   {\F}_{Selb}+\frac{1}{N_{f}^{2}}\log (c_{3}\frac{e^{-c_{2}|j|^{3}N_{f}^{2}}}{|j|^{1/8}N_{f}^{1/12}})-\frac{1}{N_{f}^{2}}\log(c_{Selb}) & \quad n^{-1}<2\sqrt{1+2\tau} \\
  \end{cases},\nonumber\\
\end{eqnarray}
where $c_1, c_2, c_3$ are defined as before.
Keeping the finite leading terms in the large $t,N_{f},N$ limit, we obtain the free energy of finite Selberg model as
\begin{eqnarray}
 \label{G free enrgy2}
 {\F}^{d}_{Selb}=  
  \begin{cases}
   \F_{Selb}  & \quad n^{-1}>2\sqrt{1+2\tau} \\
   \\
   \F_{Selb}-\frac{1}{12}\Big(\frac{n^{-1}-2\sqrt{1+2\tau}}{(1+2\tau)^{-\frac{1}{6}}\tau^{\frac{2}{3}}}\Big)^{3}   & \quad n^{-1}<2\sqrt{1+2\tau} \\
  \end{cases},
\end{eqnarray}
where $\F_{Selb}=\lim_{t,N_f\rightarrow\infty}\frac{1}{N_f^2}\sum_{j=1}^{N_f}\big(\log\Gamma(1+2t+j)+\log\Gamma(2+j)-2\log\Gamma(1+t+j)\big)$.
The phase diagram of the Selberg model is plotted in Fig.~\ref{f3}. Using the explicit formula for the free energy, Eq. \ref{G free enrgy2}, in this case, the domain wall between discrete and continuous phases in the phase diagram is of the third-order type similar to the original XX0 model.

The interpretation of the phase structure of the XX0 (Gross-Witten) model is extensively discussed in \unscite{Sa-Za}, thus for the interpretation of the Selberg model we follow the same line of argument and we declare the phase transition, like the similar one in the original model, between the discrete (finite-size) and the continuous (infinite-size) models. In the first one, the boundary size effects are present due to the access of the magnons to the boundaries of the spin chain or in other words, in this case, magnons have the freedom to move by distances of order of the size of the system. In the second case, the size of the system is infinite compared to the board of magnons diffusion and they do not reach the boundaries.

In the following, we continue with other examples of the generalized XX0 model.
\subsection*{Non-interacting XX0 spin chain; $\Delta_{m} =0$}
\label{Cons P}
As our starting case, for the sake of completeness in our mathematical study, we consider the non-interacting spin chain. 
This case corresponds to $\Delta_{m} =0$ in the Hamiltonian of the spin chain and the constant weight function in the matrix integral.
The partition function and free energy of the infinite model can be obtained from the Selberg matrix integral \ref{Selber},
\begin{equation}
\Z_{N_f}\left(f_0\right)=\prod_{j=0}^{N_f-1}\frac{\Gamma(2+j)}{\Gamma(1+j)\Gamma(2)}=M_{N_f}(0,0,1),
\end{equation}
where $f_0=const.=1$.
From the dynamical point of view, this case is a simple case with no interaction between spins and any spin only has intrinsic quantum fluctuations in its site. 
The free energy of the discrete model can be obtained by the FD formula in  Eq. \ref{Fredholm for ratio} for the ratio of the discrete to continuous model's partition functions. The asymptotic analysis of the FD and the ratio follows from the asymptotic analysis of orthogonal polynomials and in this case the analysis is simple and straightforward (Corollary 1.1 and its proof in \unscite{Baik}) and it implies the following result, 
\begin{equation}
\frac{\Z_{N_{f}}^{d}(f_0,|d|)}{\Z_{N_{f}}(f_0)}=\lim_{(N-N_f)\rightarrow\infty,N_f\rightarrow \infty}1+\mathcal{O}(e^{-c(N+N_f)}),\nonumber
\end{equation}
\begin{equation}
\label{zero potential free energy}
\F_{N_{f}}^{d}(f_0,|d|)-\F_{N_{f}}(f_0)=\lim_{N_f\rightarrow\infty}\frac{1}{N_f^2}\log(1+\mathcal{O}(e^{-c'(N_f)}))= 0 + \mathcal{O}(\frac{1}{N_f}).
\end{equation}
where $c'=c(1+n^{-1})$, $c$ is a positive real number and we used the definition of the inverse density of magnons $n^{-1} = \frac{N}{N_{f}}$.
Therefore, because of the zero interaction, the finite chain in the continuum limit always behaves like the infinite chain and as a result, there is no phase transition in this case. This is the result that one would naively guess for any continuum limit of a discrete system but as we will see in the non-zero potential, the spin chain undergoes a phase transition of order $n$ in the asymptotic limit, as $n$th derivative of the free energy discontinuously changes when passing the domain walls in different regions of the moduli space.

\subsection*{Original XX0 spin chain}
In this part, we show how to compute the free energy of the finite and infinite XX0 model near the domain walls, from the asymptotic limit of the TW distribution function.
\subsubsection*{Infinite XX0 model}
The corresponding matrix model to XX0 spin chain is the one with the Gross-Witten potential \unscite{Bogoliubov}. To obtain the free energy of infinite XX0 chain from TW, one can use the following result  which was obtained in \unscite{Baik2},
\begin{eqnarray}
\label{GW relation} 
\lim_{t\to\infty}e^{-t^{2}/4}\Z_{GW}(t)=F(x),
\end{eqnarray}
where $N_{f}=t+x(\frac{t}{2})^{\frac{1}{3}}$
and $F(x)$ is the TW distribution function \unscite{Tracy2}. 
Using the above Eq. \ref{GW relation}, the free energy of infinite XX0 model can be written as
\begin{eqnarray}
\label{free energy0} 
{\F}_{GW}=\lim_{N_{f},t\to\infty}\frac{1}{N_{f}^{2}}\Bigg(\frac{t^{2}}{4}+\log F(x)\Bigg).
\end{eqnarray}
Defining dimensionless parameters $\tau = \frac{t}{N_{f}}$ and $n^{-1} = \frac{N}{N_{f}}$, writing the argument of TW as $x=jN_{f}^{\frac{2}{3}}$ where $j=(1-\tau)/(\frac{\tau}{2})^{1/3}$, using the asymptotic results for TW distribution \unscite{Ba-Di}, which behaves as $e^{-x^\frac{3}{2}}$ in the right tail and $e^{-x^3}$ in the left tail, and implementing these in Eq.~\ref{free energy0}, we can write the free energy as
\begin{equation}
 \label{free energy0GW}
  {\F}_{GW}(\tau) = 
  \begin{cases}
   \frac{\tau^{2}}{4}+\lim_{N_{f}\to\infty}\frac{1}{N_{f}^{2}}\log (1-\frac{e^{-c_{1}j^{3/2}N_{f}}}{32\pi j^{3/2}N_{f}}) & \quad  0< \tau<1 \\
   \\
   \frac{\tau^{2}}{4}+\lim_{N_{f}\to\infty}\frac{1}{N_{f}^{2}}\log (c_{3}\frac{e^{-c_{2}|j|^{3}N_{f}^{2}}}{|j|^{1/8}N_{f}^{1/12}}) &  \quad  \tau>1\\
  \end{cases},
\end{equation}
where $c_{1} = 4/3$, $c_{2} = 1/12$ and $c_{3}=2^{\frac{1}{42}e^{(\xi^{*})'(-1)}}$.
As we see, in the right tail $x \rightarrow \infty$, we have $1-\tau>0$ and in the left tail $x \rightarrow -\infty$, we have $1-\tau<0$.
Finally, at the leading order, one can obtain
\begin{equation}
 \label{free energy0GW}
  {\F}_{GW}^{TW}(\tau) = 
  \begin{cases}
   \frac{\tau^{2}}{4} & \quad  0< \tau<1 \\
   \\
   \frac{\tau^{2}}{4}-\frac{(1-\tau)^{3}}{6\tau} &  \quad  \tau>1\\
  \end{cases}.
\end{equation}
The above expression gives the free energy in the vicinity of the domain wall $\tau=1$. It also shows a third-order phase transition in XX0 spin chain, Fig.~\ref{f1}.
Moreover, this is interesting to compare the free energy \ref{free energy0GW} with the original result previously obtained from other plausible methods \unscite{Gross},
\begin{equation}
 \label{free energy0GWorgin}
  {\F}_{GW}(\tau) = 
  \begin{cases}
   \frac{\tau^{2}}{4} & \quad  0< \tau<1 \\
   \\
   \tau-\frac{3}{4}-\frac{\log{\tau}}{2} &  \quad  \tau>1\\
  \end{cases},
\end{equation}
and as we expect TW free energy in Eq. \ref{free energy0GW} and its derivative up to third order coincide and agree with the Gross-Witten results in Eq. \ref{free energy0GWorgin} in the vicinity of the domain wall $\tau=1$. This is another clue for our proposed conjecture, that the TW contains all the asymptotic information of the model near the domain walls of the phase transitions.

\subsubsection*{Finite XX0 model}

Using the asymptotic analysis of FD \unscite{Baik} and Eq.~\ref{free energy0GWorgin}, the free energy of finite XX0 model, near the domain walls, is obtained in our previous study \unscite{Sa-Za},
\begin{equation}
 \label{free enrgy1}
 \F_{GW}^{d}=  
  \begin{cases}
   \frac{\tau^{2}}{4}  & \quad n^{-1} > \tau+1 \quad \& \quad\tau \leq 1\\
   \\
   \frac{\tau^{2}}{4}-\frac{c_{2}}{2^{-1}\tau}|n^{-1}-(\tau+1)|^{3}  &  \quad n^{-1} < \tau+1\quad \& \quad\tau \leq 1\\
   \\
   \tau-\frac{{3}}{4}-\frac{{\log \tau}}{2} & \quad n^{-1} > 2\sqrt \tau\quad \& \quad\tau > 1 \\
   \\
   \tau-\frac{{3}}{4}-\frac{{\log \tau}}{2} -\frac{c_{2}}{2^{-2}\tau(\tau^{\frac{1}{2}}+\tau^{-\frac{1}{2}})}|n^{-1}-2\sqrt \tau|^{3} & \quad n^{-1} < 2\sqrt \tau\quad \& \quad\tau > 1\\   
  \end{cases}.
\end{equation}
The phase structure of the finite XX0 chain is plotted in Fig.~\ref{f2}, and it is easy to see from Eq. \ref{free enrgy1} that the solid line is of the second-order transition type and the other domain walls are of the third order transition type.

\subsubsection*{Weakly coupled XX0 model; $\Delta_m=c\delta_{m,1}, c << 1$}
As we explained in \unscite{Sa-Za}, a possible interpretation of the quadratic matrix model is the weakly coupled XX0 model. The suitable potential for this case is quadratic potential which is the leading term of Gross-Witten potential. In the matrix integral framework, this potential is corresponding to a short-range interaction XX0 model with weak coupling constant. In contrast to GW model, this system, in the infinite scale, do not experience any phase transition. However, due to the FD relation for the discrete-to-continuous ratio, the free energy of finite XX0 with weak coupling constant jumps across the domain wall as, \unscite{Sa-Za},  
\begin{eqnarray}
 \label{Quad free enrgy}
 \F^{d}_{Quad}=  
  \begin{cases}
   \mathcal{\F}_{Quad}  & \quad \lambda > 2 \\
   \\
   \mathcal{\F}_{Quad}-\frac{1}{3}|\lambda-2|^{3}   & \quad \lambda < 2 \\
  \end{cases},
\end{eqnarray}
where $\mathcal{\F}_{Quad}=\lim_{N_f\rightarrow\infty}\frac{1}{N_f^2}\sum_{j=1}^{N_f}\left(\log\Gamma(1+j)-\log\Gamma(2)\right)$ is the free energy of the infinite model obtained from Mehta integral \unscite{Mehta}, and $\lambda=N/\sqrt{N_f}$ is the fixed parameter. According to the above free energy \ref{Quad free enrgy}, the spin chain experiences a third-order discrete-to-continuous phase transition. 

\section{Conclusion and discussion}
In this work, we studied the phase structure of the generalized XX0 model in concrete examples, including the infinite-range interacting model, namely the Selberg model. We observed, similar to the original model, a discrete-to-continuous third-order phase transition occurs in the asymptotic limit of this model. This is expected that the similar phase transitions happen in the generalized XX0 model with any form of the potential. As we conjecture, all these phase transitions can be explained by TW and the phase structure of the generalized model contains similar phase transitions as well as new phase transitions.
The actual mathematical proof of the universality and more details about the complete phase structure of the infinite/finite generalized model and its interpretations will appear in a separate work \unscite{Zah}.

Our universality conjecture is formulated as follows. In the generalized XX0 model and its examples with any form of the potential, always there is a third-order phase transition between finite- and infinite-size (discrete/continuous) model which is governed by the TW. For the finite-length interacting generalized XX0 model, always there is a Gross-Witten type phase transition which is either of third-order or second-order, depending on the type of the domain wall. If the GW-type transition is between two infinite models this is of third-order and if the transition is between two finite-size models, this is of second-order. In fact, since the phase transition is hidden in the mathematical structure of the TW in the asymptotic limit an this is possible to obtain the free energy of the model in the vicinity of the domain walls from the TW distribution, thus we conjecture that all the phase transitions in the model are governed by the TW distribution. The existence of the TW distribution in the asymptotic limit of the partition function is always associated with a phase transition of the second- or third-order type. For a recent similar study, see the review \unscite{Sc-Ma}.

We conjecture that the obtained results in chapter 3, for the specific examples of the generalized XX0 model, in terms of the TD and their asymptotic limits, TW, can be generalized to any model in the class of generalized XX0 model. In fact, the TW distribution appears in the asymptotic limits of the partition function of the  generalized XX0 model and its finite/infinite examples as the following formal expression,
\begin{eqnarray}
\label{Z/Z=F} 
\lim_{min(N,N_{f},t)\to\infty}\frac{\Z_{N_f}^{d}(f)}{\Z_{N_f}(f)}=F(x),\quad \lim_{min(N_{f},t)\to\infty}\Z_{N_f}(f')=F(y),
\end{eqnarray}
where $f$ is the weight function of general potential, Eq. \ref{V}, or any example in this class, and $f'$ is the one for short/long, but excluding infinite-range potential, and the arguments of the TW distribution $F$, $x$ and $y$ have to be determined for the general model and its specific cases from the microscopic details of the system such as the exact form of the coupling constant in the Hamiltonian. The left expression in Eq. \ref{Z/Z=F} determines the discrete/continuous phase transition whereas the right one determines the GW-type phase transition. In fact, the argument characterizes the domain wall in the phase diagram. The arguments can always be written in terms of the parameters of the model as
\begin{equation}
x=j(\tau,n^{-1})N_{f}^{\frac{2}{3}},\quad y=j(\tau)N_{f}^{\frac{2}{3}},
\end{equation}
where $j(\tau,n^{-1})$ and $j(\tau)$ are dimensionless functions and their forms depend on the model. The domain wall in each type of phase transition is, in fact, the nominator of the function $j$.
The appearance of the TW distribution in the partition function of the model and the fact that the TW distribution is asymmetric, i.e. the left and right tails are different asymptotically, always lead to the phase transitions in the generalized XX0 model.

Our heuristic argument for the universality consists of two interconnected parts. First, it is a common understanding in critical phenomena, that for the models in a universality class, in our case the class of generalized XX0 model, the phase structure, i.e. the regions, domain walls and their orders in the phase space, is independent of the microscopic interactions and potentials. In fact, the phase transition is a global, large scale collective behavior of the system in which the microscopic details of the system are negligible and washed out \unscite{Gold}. Second, all the phase transitions in the generalized XX0 model are governed by the TW distribution. In RMT, the universality of the TW distribution with respect to the form of the potentials has been conjectured and it is believed that the fluctuations of the systems around their boundaries in the asymptotic limit are distributed and controlled by TW distribution \unscite{Dei}. The relation between the universality of TW distribution and the universality in critical phenomena and phase transition is heuristically discussed in the review \unscite{Sc-Ma}. Thus, it is natural for the generalized XX0 model, which is a matrix model, to be described in the asymptotic by the TW distribution. The precise relations between TW universality and the phase transitions in matrix models will be discussed in our future work.

One of our immediate goals for the extension of this study is to apply the Riemann-Hilbert problem in the generalized XX0 model to prove the universality conjecture, rigorously. Using the well-developed techniques of asymptotic analysis of FD, the phase structure of the generalized model can be studied and the universal features can be extracted.

In this study, we have partially explored the phase structure of the generalized XX0 model, in some examples. The complete phase structure of the generalized model is a motivated natural extension of this work. From the matrix model perspective of the general model \unscite{Ju-Za}, it is apparent that the phase structure of the generalized XX0 model is a rich sophisticated structure and there are other third- and second-order phase transitions due to the competitions between different $\Delta_m$'s. Understanding and interpreting the phase structure of the generalized XX0 model using the TW technique and its comparison with the known results \unscite{Ju-Za} is one of the main goals of our future studies.

From another plausible approach to the phase structure, the explicit free energy of the generalized XX0 model and its specific cases can be obtained by applying the standard methods of random matrix theory. In fact, case by case by inserting the (continuous and discrete) eigenvalue density $\rho$, which is the minimizer of the following action formula, one can obtain the free energy.
We plan to explicitly obtain the free energy and phase structure of these models. Then, we can compare and check the results obtained from the new techniques of the TW with the results from the above standard method.  
In the nearest and next-to-nearest interacting XX0 model which is another special case of the generalized model, the phase structure of the model is studied in detail in \unscite{Ju-Za}. We translate these results in spin chain language and interpret them following the introduced interpretation in \unscite{Sa-Za}. 
These are the results for the Gross-Witten type phase transitions in the infinite generalized model.

The class of generalized spin chain models that is studied in this article and the conjecture about their universal phase structure are closely related to the universality in random matrix theory.
However, there are variants of Gross-Witten matrix models called multi-critical matrix model with the multi-critical potential, i.e. a higher order polynomial potentials. As studied extensively, the multi-critical models have different critical behavior \unscite{Pe-Sh 1}, \unscite{Pe-Sh 2}. Recently, these models are discovered in the context of the non-interacting fermions in anharmonic potential, \unscite{Le-Ma}. The multi-critical matrix models are also discussed in the context of XXZ models, \unscite{Ste}. In principle, these models and potentials can be written in terms of the generalized Gross-Witten potential with several nonzero $\Delta_m$'s by tuning the coefficients. These matrix models have more complicated phase structure which are explained by the generalization of the Tracy-Widom distribution \unscite{Cl-Kr}, \unscite{Ak-At}. The study of these models and their phase structures is an interesting line of research for future.

Besides the zero anisotropy limit of the XXZ model (XX0 model) which is the main topic of this study, there is an interesting and closely related limit, namely the infinite anisotropy limit ($\Delta_z\rightarrow -\infty$), called the Ising limit. In this limit, the Hamiltonian of the spin chain is given by the same Hamiltonian as XX0 model except with an insertion of the projection operator which annihilates any state with down spin in the neighbouring sites, \unscite{Ab-Ig}, \unscite{Bo-Ma 2}. The spin chain described by this Hamiltonian is called effective model.
As a result of study in \unscite{Bo-Ma rev}, the effective model, similar to the original XX0 model, under
appropriate limiting conditions leads to the
Gross-Witten partition function. Therefore, it is natural to expect that the asymptotic behavior and phase structure for XXZ model in the Ising limit is similar to those of XX0 model. The similarity in the properties of both models are possibly originated in the same combinatorial structure for both models, \unscite{Bo-Ma com}.


Furthermore, stemming from the analogy with the strongly coupled bosonic system, \unscite{Bo-Iz}, the effect of the projection operator can be seen as a coupling of second-order and a quasi-local interaction beyond the nearest-neighbour. These higher (beyond first)-order effects in spin chain motivates further studies on the possibility of a rigorous relation between the effective model and the generalized XX0 model.

Qualitatively speaking, the class of universality in the equilibrium phase transitions are sensitive to the symmetries of the models rather than microscopic interactions. In the asymptotic limit, both XX0 and effective model are described by the Gross-Witten matrix model because of the same symmetries at the level of combinatroics, \unscite{Bo-Ma com}. The generalized XX0 model and the effective model are contrasted against XX0 model by the length of the interaction and thus, we conjecture that the generalized XX0 models as well as effective model experience the Gross-Witten transition as XX0 does. 
All in all, as an interesting extension of the present work, we propose a detailed study of the apparent connections between generalized XX0 model and the effective model, both at finite parameters and in the asymptotic limit to understand the phase structure of the effective model. 

\section{Acknowledgement}
We are grateful to Prof. Peter Forrester and Prof. Ole Warnaar for their correspondences. A.Z. appreciates the research funds from National Institute for Theoretical Physics,
School of Physics and Mandelstam Institute for Theoretical Physics,
University of the Witwatersrand. A.Z. is deeply grateful to Prof. Robert de Mello Koch for his support during the preparation of this work. The authors deeply appreciate Ezad Shojaei, without whom this study would not have been initiated. M. S. acknowledge financial support from Ministerio de Econom{\'\i}a y Competitividad (MINECO) and Fondo Europeo de Desarrollo Regional (FEDER) under project ESOTECOS FIS2015-63628-C2-2-R.

\appendix
\section{Non-intersecting brownian motion in selberg potential}
\label{MIG}
In this section, we review the known results regarding the width of the foremost walkers of NIBM in the Selberg potential and its asymptotic limits, the TW distribution obtained in \unscite{Baik}.
These results are one of starting points in our study to compute the free energy of the Selberg model and extract the phase structure. 

Consider  $X(t'):=(X_0(t'), X_1(t'), ..., X_{N_f-1}(t'))$ as independent discrete-time simple symmetric random walk. The distance of these nonintersecting walkers away from origin are defined as $X_{i}(t'), i=0,...,N_f-1$ and they are subject to initial condition, $X(0):=( 0 , 2 ,..., 2N_{f}-2)$ and final condition, $X(0)=X(2t)$, and the following condition from the nonintersecting character of the process
$X_{0}(t')<X_{1}(t')<...<X_{N_{f}-1}(t')$ for all $t'=0,1,...,2t$. The maximum distance between first and last walker is defined as the width $W_{N_{f}}(2t)=max_{t'=0,1,...,2t}\big(X_{N_{f}-1}(t')-X_{0}(t')\big)$. It has been proved in \unscite{Baik} that the conditional probability on the width in the domain $d_s$ with size $|d_s|=N$; $d_{s}=\{z\in \mathds{C} |z^{N}=s\}$, can be expressed in terms of TD as,
\begin{eqnarray}
\label{con prob1}
\mathbb{P}(W_{N_{f}}(2t)<2N) = \oint_{|s|=1} \frac{D_{N_{f}}^{d_{s}}(f_{Selb},|d_{s}|)}{D_{N_{f}}(f_{Selb})} \frac{ds}{2\pi i s}, \hspace{.5cm} f_{Selb}(z)=z^{-t}(1+z)^{2t}.
\end{eqnarray}
The width of the process always satisfies $2N_{f}\leq W_{N_{f}}(2t)\leq 2N_{f}+2t$ and for any fixed value of $\tau=\frac{t}{N_f}>0$, the fluctuations are given by TW as,
\begin{eqnarray}
\label{con prob2}
\lim_{t\to\infty}\mathbb{P}\left(\frac{W_{N_{f}}(2t)-2\sqrt{(N_{f}^{2}+2N_{f}t)}}{(N_{f}^{2}+2N_{f}t)^{-\frac{1}{6}}t^{\frac{2}{3}}}\leqslant x\right)=F(x),
\end{eqnarray}
for each x $\in$ $\mathbb{R}$. This result is more general than our physical case of spin chain with parameter $\tau$ and in fact it is valid for any fixed $\gamma=\frac{N_f}{t^\beta}>0$ and $0<\beta<2$. 

\section{Expansion of the Selberg integral}
\label{APP G P}
In this appendix, we present the details of the expansion of the Selberg potential in the form of the general potential \ref{V}. 
One possible infinitely long-range potential for the generalized XX0 Hamiltonian \ref{General H} is the Selberg potential defined by 
\begin{eqnarray}
\label{General1} 
V(z)&=&\frac{1}{t}\log\big(z^{-t}(1+z)^{2t}\big)=\frac{1}{t}\log\big(z^{-t}(1+2z+z^{2})^{t}\big)=\frac{1}{t}\log\big((z^{-1}+2+z)^{t}\big)\nonumber\\
&=&\log\big(2+z^{-1}+z\big)=\log\big(1+\frac{z^{-1}+z}{2}\big)+\log 2\nonumber\\
&=&\sum_{j=1}^{\infty}\frac{(-1)^{j-1}}{j2^{j}}(z^{-1}+z)^{j} + \log 2 \nonumber\\
&=&\sum_{j=1}^{\infty}B_j(z^{-1}+z)^{j} + \log 2 \nonumber\\
&=&\lim_{N\to\infty}\sum_{j=1}^{N}J_{j} (z^{-j}+z^{j}) +\log 2,
\end{eqnarray}
where $B_j=\frac{(-1)^{j-1}}{j2^{j}}$ is defined in the forth line and $J_{j}$ is the coefficients of the expansion to be determined in Eq. \ref{JDelta}.
To find the coefficients $J_j$, one needs to evaluate a sum over binomial expansion of,
\begin{eqnarray}\nonumber
(z+z^{-1})^1&=&z^1+z^{-1} \\ \nonumber
(z+z^{-1})^2&=&z^{2}+z^{-2}+\binom{2}{1}  \\ \nonumber
(z+z^{-1})^3&=&z^{3}+z^{-3}+\binom{3}{1}z^{1}+\binom{3}{2}z^{-1}  \\ \nonumber
(z+z^{-1})^4&=&z^{4}+z^{-4}+\binom{4}{1}z^{2}+\binom{4}{3}z^{-2}+\binom{4}{2}  \\ \nonumber
(z+z^{-1})^5&=&z^{5}+z^{-5}+\binom{5}{1}z^{3}+\binom{5}{4}z^{-3}+\binom{5}{2}z^{1}+\binom{5}{3}z^{-1}  \\ \
(z+z^{-1})^6&=&z^{6}+z^{-6}+\binom{6}{1}z^{4}+\binom{6}{5}z^{-4}+\binom{6}{2}z^{2}+\binom{6}{4}z^{-2}+\binom{6}{3}  \\ \nonumber
(z+z^{-1})^7&=&z^{7}+z^{-7}+\binom{7}{1}z^{5}+\binom{7}{6}z^{-5}+\binom{7}{2}z^{3}+\binom{7}{5}z^{-3}+\binom{7}{3}z^{1}+\binom{7}{4}z^{-1}  \\ \nonumber
\vdots \\ \nonumber
(z+z^{-1})^j&=&\sum_{k=0}^{j}\binom{j}{k}z^{j-2k}\nonumber\\
&=&z^j+z^{-j}+\binom{j}{1}z^{j-2}+\binom{j}{j-1}z^{-(j-2)}+\binom{j}{2}z^{j-4}+\binom{j}{j-2}z^{-(j-4)}+\cdots.\nonumber\\
\end{eqnarray}
It is easy to see the general pattern for the sum from an example up to some order, thus, one can recognize the following pattern for the general sum,
\begin{eqnarray}
B_{1}(z+z^{-1})^1+B_{2}(z+z^{-1})^2+\cdots+B_{N}(z+z^{-1})^N&=&\sum_{j=1}^{N}\Bigg(\sum_{i=0}^{[\frac{N-j}{2}]}B_{2i+j}\binom{2i+j}{i}z^{j}+B_{2i+j}\binom{2i+j}{2i+j-i}z^{-j}\Bigg)\nonumber \\
&=&\sum_{j=1}^{N}\Bigg(\sum_{i=0}^{[\frac{N-j}{2}]}B_{2i+j}\binom{2i+j}{i}(z^{j}+z^{-j})\Bigg).
\end{eqnarray}
Therefore, we obtain the coefficient $J_j$ and the coupling $\Delta_m$ in the expansion of Selberg potential in the form of the general potential \ref{V} as
\begin{equation}
\label{JDelta}
J_{j}=\sum_{i=0}^{\big[\frac{N-j}{2}\big]}B_{2i+j}\binom{2i+j}{i},\quad \Delta_m= \sum_{i=0}^{\big[\frac{N-m}{2}\big]}B_{2i+m}\binom{2i+m}{i}. 
\end{equation}

\section{Matrix integrals and observables of generalized XX0 model}
\subsection{Selberg integrals and factorization of symmetric functions}
\label{schur function}
Let us review two mathematical results that are essential in the calculations of the correlation functions of generalized XX0 model. The Selberg integral is a generalized form of Euler beta function in any dimension, (see \unscite{Forrester1} for a review),
\begin{eqnarray}
\label{Selber}
M_{N_f}(a,b,\gamma)&\coloneqq&\frac{1}{(2\pi)^{N_f}}\int_{-\pi}^{\pi}...\int_{-\pi}^{\pi}\prod_{i=0}^{N_f}e^{\frac{1}{2}\imath\theta_i(a-b)}|1+e^{\imath\theta_i}|^{a+b}
\prod_{1\leq i< j\leq N_f}|e^{\imath\theta_i}-e^{\imath\theta_j}|^{2\gamma}d\theta_1...d\theta_{N_f}\nonumber \\
&=&\prod_{j=0}^{N_f-1}\frac{\Gamma(1+a+b+j\gamma)\Gamma(1+(j+1)\gamma)}{\Gamma(1+a+j\gamma)\Gamma(1+b+j\gamma)\Gamma(1+\gamma)},
\end{eqnarray}
where $Re(a + b + 1) > 0, \quad Re(\gamma) > - min\{1/n,\quad Re(a + b + 1)/(n - 1)\}$. The Selberg integral in this form is the general formula that its specific cases will be used in the computations of the partition functions and correlation functions.

The Jack polynomials are a class of multivariate orthogonal polynomials denoted by $J_{\lambda}^{(1/\gamma)}$. They are one-parameter $\gamma$ generalization of the Schur polynomials \unscite{Forrester1}, i.e. $J_{\lambda}^{(1/\gamma)}|_{\gamma=1}=S_{\lambda}$. The factorization property of one Jack polynomial is expressed as
\begin{eqnarray}
\label{Jack}
\frac{1}{(2\pi)^{N_f}}\int_{-\pi}^{\pi}&...&\int_{-\pi}^{\pi}J_{\lambda}^{(1/\gamma)}(-e^{\imath\theta})\prod_{i=0}^{N_f}e^{\frac{1}{2}\imath\theta_i(a-b)}|1+e^{\imath\theta_i}|^{a+b}
\prod_{1\leq i< j\leq N_f}|e^{\imath\theta_i}-e^{\imath\theta_j}|^{2\gamma}d\theta_1...d\theta_{N_f} \\ \nonumber
&=&\frac{[-b]_{\lambda}^{(\gamma)}}{[1+a+(n-1)\gamma]_{\lambda}^{(\gamma)}}J_{\lambda}^{(1/\gamma)}(1^{N_f})M_{N_f}(a,b,\gamma),
\end{eqnarray}
where $[b]_{\lambda}^{(\gamma)} =\prod_{i\geq1}(b + (1 - i)\gamma)_{\lambda_i}$ with $(b)_n=b(b+1)...(b+n-1)$ a Pochhammer symbol \unscite{Forrester1}. With two Jack polynomials inserted in the Selberg integral, up to a numerical factor, the following result for the special case $a+b=0$ is obtained \unscite{Kad},
\begin{eqnarray}
\label{Two Jacks}
\frac{1}{(2\pi)^{N_f}}\int_{-\pi}^{\pi}&...&\int_{-\pi}^{\pi}J_{\mu}^{(1/\gamma)}(-e^{-\imath\theta})J_{\lambda}^{(1/\gamma)}(-e^{\imath\theta})\prod_{i=0}^{N_f}e^{\imath\theta_i a}
\prod_{1\leq i< j\leq N_f}|e^{\imath\theta_i}-e^{\imath\theta_j}|^{2\gamma}d\theta_1...d\theta_{N_f}\nonumber \\
&=&J_{\mu}^{(1/\gamma)}(1^{N_f})J_{\lambda}^{(1/\gamma)}(1^{N_f})\prod_{i=1}^{N_f}\frac{(\gamma-1)!}{(\gamma-a-1-\lambda_i+\mu_i)!(a+\lambda_i-\mu_i)!}\nonumber\\
&\times&\prod_{1\leq i,j\leq N_f}\frac{1}{(1+a-\gamma+(j-i)\gamma+\lambda_i-\mu_j)_{\gamma}(-a+(j-i)\gamma-\lambda_i+\mu_j)_{\gamma}}.
\end{eqnarray}
As a special case $\gamma=1$ of Eq. \ref{Jack}, 

\begin{eqnarray}
\label{Jack0}
<S_{\lambda}(-e^{i\theta})>&\coloneqq&\frac{1}{(2\pi)^{N_f}}\prod_{i=1}^{N_f}\int_{-\pi}^{\pi}d\theta_i S_{\lambda}(-e^{\imath\theta})e^{\frac{1}{2}\imath\theta_i(a-b)}|1+e^{\imath\theta_i}|^{a+b}\prod_{1\leq i< j\leq N_f}|e^{\imath\theta_i}-e^{\imath\theta_j}|^{2}\nonumber \\
&=&\frac{[-b]_{\lambda}}{[1+a+(n-1)]_{\lambda}}S_{\lambda}(1^{N_f})\prod_{j=0}^{N_f-1}\frac{\Gamma(1+a+b+j)\Gamma(2+j)}{\Gamma(1+a+j)\Gamma(1+b+j)\Gamma(2)}\nonumber \\
&=&\frac{[-b]_{\lambda}}{[1+a+(n-1)]_{\lambda}}S_{\lambda}(1^{N_f})M_{N_f}(a,b),
\end{eqnarray}
where $[b]_{\lambda} =\prod_{i\geq1}(b + (1 - i))_{\lambda_i}$, and $Re(a + b + 1) > 0$.
The case of two Schur functions, the special case $\gamma=1$ of Eq. \ref{Two Jacks}, can be evaluated for $a+b=0$, using Theorem 5 in \unscite{Kad}, up to a numerical factor, as
\begin{eqnarray}
\label{Jack00}
&&<S_{\mu}(-e^{-i\theta})S_{\lambda}(-e^{i\theta})>\nonumber\\
&\coloneqq&\frac{1}{(2\pi)^{N_f}}\prod_{i=1}^{N_f}\int_{-\pi}^{\pi}d\theta_iS_{\mu}(-e^{-\imath\theta})S_{\lambda}(-e^{\imath\theta})e^{\frac{1}{2}\imath\theta_i(a-b)}|1+e^{\imath\theta_i}|^{a+b} \prod_{1\leq i< j\leq N_f}|e^{\imath\theta_i}-e^{\imath\theta_j}|^{2}\nonumber\\
&=&S_{\mu}(1^{N_f})S_{\lambda}(1^{N_f})\prod_{i=1}^{N_f}\frac{1}{(a+\lambda_i-\mu_i)!^2}\prod_{1\leq i,j\leq N_f}\frac{1}{(j-i)^2-(a+\lambda_i-\mu_j)^2}.
\end{eqnarray}
For more convenience let us define $K_{N_f}(a;\mu,\lambda)\coloneqq\prod_{i=1}^{N_f}\frac{1}{(a+\lambda_i-\mu_i)!^2}\prod_{1\leq i,j\leq N_f}\frac{1}{(j-i)^2-(a+\lambda_i-\mu_j)^2}$.
\subsection{Partition functions and correlations}
\label{C_P}
In the following, we present explicit calculations of the observables (partition functions and correlation function) of the generalized infinite-size XX0 model and in its limiting cases, Selberg, Gross-Witten, quadratic and single-term models. First, we start with the partition functions, and by expanding the potential and finding its expansion form as a specification of the Selberg integral, we obtain explicit expressions for the partition functions. Second, using the factorization property of Schur functions in the Selberg matrix integrals, we calculate the correlation functions. The large size limit ($N\rightarrow \infty$) is implicitly assumed for all the formulas in this section.

For computing the partition and correlation functions in the case of the general model and its examples, quadratic, Gross-Witten, and single-term models, we prescribe the following procedure. Let fix \textit{a} and \text{b} such that the weight function in Eq.~\ref{Jack0}, $e^{\frac{1}{2}\imath\theta_i(a-b)}|1+e^{\imath\theta_i}|^{a+b}$, became $z^{a}$ where $z=e^{\imath \theta}$. Besides, we can expand the general, quadratic, Gross-Witten and single-term potentials in terms of $z^{a}$. Thus, by a simple summation over Eq.~\ref{Jack0} we can obtain the partition and correlation functions of the general, quadratic, Gross-Witten and single-term models.
In these cases, we need to manipulate Eq.~\ref{Jack} such that the desired potential appears in the left side of the identity. 
As an starting example, we set $a+b=0$ to provide a following form of potential in a typical correlation function,   
\begin{eqnarray}
\label{Jack1}
\frac{1}{(2\pi)^{N_f}}\int_{-1}^{1}...\int_{-1}^{1}dz_iS_{\lambda}(z)\prod_{i=1}^{N_f}z_i^{a-1}\prod_{1\leq i< j\leq N_f}|z_i-z_j|^2\\ \nonumber
=\frac{(-1)^{-|\lambda|}[a-1]_{\lambda}}{[a-1+N_f]_{\lambda}}S_{\lambda}(1^{N_f})M_{N_f}(a-1,1-a,1).
\end{eqnarray}
Now, we make summation over $z_i^{a}$ such that the form of quadratic potential appears.
Thus, the following summation over Selberg integral gives the partition function of XX0 model with weak coupling constant and
it can be expressed as,
\begin{eqnarray}
\label{Jack2Z}
\Z_{Quad}&=&\bra{\Uparrow}\prod_{j=1}^{N_f}\sigma^{+}_{N_f-j}e^{-t{\hat{H}}_{Quad}}\prod_{i=1}^{N_f}\sigma^{-}_{N_f-i}\ket{\Uparrow}\nonumber\\
&=&\frac{1}{(2\pi)^{N_f}}\prod_{i=1}^{N_f}\int_{-1}^{1}dz_i\Bigg(\sum_{a}^{\infty}\frac{(-1)^{2a}t^az_{i}^{2a}}{2^aa!}\Bigg) \prod_{1\leq i< j\leq N_f}|z_i-z_j|^2\nonumber\\
&=&\sum_{a=0}^{\infty}\frac{(-1)^{2a}t^a}{2^a a!}M_{N_f}(2a,-2a,1).
\end{eqnarray}
The special correlation function of the quadratic model can be similarly computed,
\begin{eqnarray}
\C_{Quad}^*&=&\bra{\Uparrow}\prod_{i=1}^{N_f}\sigma^{+}_{N_f-j}e^{-t{\hat{H}}_{Quad}}\prod_{i=1}^{N_f}\sigma^{-}_{l_i}\ket{\Uparrow}\nonumber\\
&=&\frac{1}{(2\pi)^{N_f}}\prod_{i=1}^{N_f}\int_{-1}^{1}dz_iS_{\lambda}(z)\Bigg(\sum_{a}^{\infty}\frac{(-1)^{2a}t^az_{i}^{2a}}{2^aa!}\Bigg)\prod_{1\leq i< j\leq N_f}|z_i-z_j|^2\nonumber \\
&=&(-1)^{-|\lambda|}S_{\lambda}(1^{N_f})\sum_{a=0}^{\infty}\frac{(-1)^{2a}t^a}{2^a a!}\frac{[2a]_{\lambda}}{[2a+N_f]_{\lambda}}M_{N_f}(2a,-2a,1).
\end{eqnarray}
Moreover, using the current techniques \unscite{Kad}, the full correlation function of the quadratic model can be calculated, up to a numerical factor, as
\begin{eqnarray}
\C_{Quad}&=&\bra{\Uparrow}\prod_{i=1}^{N_f}\sigma^{+}_{j_i}e^{-t{\hat{H}}_{Quad}}\prod_{i=1}^{N_f}\sigma^{-}_{l_i}\ket{\Uparrow}\nonumber\\
&=&\frac{1}{(2\pi)^{N_f}}\prod_{i=1}^{N_f}\int_{-1}^{1}dz_iS_{\mu}(z^{-1})S_{\lambda}(z)\Bigg(\sum_{a}^{\infty}\frac{(-1)^{2a}t^az_{i}^{2a}}{2^aa!}\Bigg)\prod_{1\leq i< j\leq N_f}|z_i-z_j|^2\nonumber \\
&=&(-1)^{|\mu|-|\lambda|}S_{\mu}(1^{N_f})S_{\lambda}(1^{N_f})\sum_{a=0}^{\infty}\frac{(-1)^{2a}t^a}{2^a a!}K_{N_f}(2a; \mu, \lambda).
\end{eqnarray}

The same procedure applies for computing the matrix integral representation of the partition function and correlation functions of the Gross-Witten model. By using the expansion of the GW potential as,
\begin{equation}
f_{GW} =e^{t(z+z^{-1})}\\
=\sum_{j=0}^{\infty}\frac{t^j}{j!}(z+z^{-1})^j\\
=\sum_{j=0}^{\infty}L_j(z^{j}+z^{-j}),
\end{equation}
where $L_{j}=\sum_{i=0}^{\big[\frac{N-j}{2}\big]}\frac{t^{(2i+j)}}{(2i+j)!}\binom{2i+j}{i}$, see Appendix B for the similar analysis. The partition function of the Gross-Witten XX0 model can be written in terms of the Selberg integrals as,
\begin{eqnarray}
\label{Jack2ZZ}
\Z_{GW}&=&\bra{\Uparrow}\prod_{j=1}^{N_f}\sigma^{+}_{N_f-j}e^{-t{\hat{H}}_{GW}}\prod_{i=1}^{N_f}\sigma^{-}_{N_f-i}\ket{\Uparrow}\nonumber\\
&=&\frac{1}{(2\pi)^{N_f}}\prod_{i=1}^{N_f}\int_{-1}^{1}dz_i\Bigg(\sum_{a=0}^{\infty}L_a(z_i^{a}+z_i^{-a})\Bigg) \prod_{1\leq i< j\leq N_f}|z_i-z_j|^2\nonumber\\ 
&=&\sum_{a=0}^{\infty}L_a\Big(M_{N_f}(a,-a,1)+M_{N_f}(-a,a,1)\Big).
\end{eqnarray}
The special correlation function of Gross-Witten model can be computed as
\begin{eqnarray}
\C_{GW}^*&=&\bra{\Uparrow}\prod_{i=1}^{N_f}\sigma^{+}_{N_f-j}e^{-t{\hat{H}}_{GW}}\prod_{i=1}^{N_f}\sigma^{-}_{l_i}\ket{\Uparrow}\nonumber\\
&=&\frac{1}{(2\pi)^{N_f}}\prod_{i=1}^{N_f}\int_{-1}^{1}dz_iS_{\lambda}(z)\Bigg(\sum_{a=0}^{\infty}L_a(z_i^{a}+z_i^{-a})\Bigg)\prod_{1\leq i< j\leq N_f}|z_i-z_j|^2\nonumber \end{eqnarray}
\begin{eqnarray}
=(-1)^{-|\lambda|}S_{\lambda}(1^{N_f})\sum_{a=0}^{\infty}L_a\Big(\frac{[a]_{\lambda}}{[a+N_f]_{\lambda}}M_{N_f}(a,-a,1)+\frac{[-a]_{\lambda}}{[-a+N_f]_{\lambda}}M_{N_f}(-a,a,1)\Big).\nonumber\\
\end{eqnarray}
Similarly, the full correlation in this model can be calculated, up to a numerical factor, as
\begin{eqnarray}
\C_{GW}&=&\bra{\Uparrow}\prod_{i=1}^{N_f}\sigma^{+}_{j_i}e^{-t{\hat{H}}_{GW}}\prod_{i=1}^{N_f}\sigma^{-}_{l_i}\ket{\Uparrow}\nonumber\\
&=&\frac{1}{(2\pi)^{N_f}}\prod_{i=1}^{N_f}\int_{-1}^{1}dz_iS_{\mu}(z^{-1})S_{\lambda}(z)\Bigg(\sum_{a=0}^{\infty}L_a(z_i^{a}+z_i^{-a})\Bigg)\prod_{1\leq i< j\leq N_f}|z_i-z_j|^2\nonumber\\
&=&(-1)^{|\mu|-|\lambda|}S_{\mu}(1^{N_f})S_{\lambda}(1^{N_f})\sum_{a=0}^{\infty}L_a\Big(K_{N_f}(a; \mu, \lambda)+K_{N_f}(-a; \mu, \lambda)\Big).\nonumber\\
\end{eqnarray}

In the first level of generalization of the Gross-Witten case, consider the interaction with the $n$th neighbours $V=\Delta_n (z^n+z^{-n})$, then the partition function can be obtained as
\begin{eqnarray}
\label{Jack2ZZ}
\Z_{n}&=&\bra{\Uparrow}\prod_{j=1}^{N_f}\sigma^{+}_{N_f-j}e^{-t{\hat{H}}_{n}}\prod_{i=1}^{N_f}\sigma^{-}_{N_f-i}\ket{\Uparrow}\nonumber\\
&=&\frac{1}{(2\pi)^{N_f}}\prod_{i=1}^{N_f}\int_{-1}^{1}dz_i\Bigg(\sum_{a=0}^{\infty}L_{n,a}(z_i^{na}+z_i^{-na})\Bigg) \prod_{1\leq i< j\leq N_f}|z_i-z_j|^2\nonumber\\ 
&=&\sum_{a=0}^{\infty}L_{n,a}\Big(M_{N_f}(na,-na,1)+M_{N_f}(-na,na,1)\Big),
\end{eqnarray}
where $L_{n,a}=\sum_{i=0}^{\big[\frac{N-a}{2}\big]}\frac{t^{(2i+a)} (\Delta_n)^{2i+a}}{(2i+a)!}\binom{2i+a}{i}$ . Similarly for the special correlation functions we have, 
\begin{eqnarray}
\C_{n}^*&=&\bra{\Uparrow}\prod_{i=1}^{N_f}\sigma^{+}_{N_f-j}e^{-t{\hat{H}}_{n}}\prod_{i=1}^{N_f}\sigma^{-}_{l_i}\ket{\Uparrow}\nonumber\\
&=&\frac{1}{(2\pi)^{N_f}}\prod_{i=1}^{N_f}\int_{-1}^{1}dz_iS_{\lambda}(z)\Bigg(\sum_{a=0}^{\infty}L_{n,a}(z_i^{na}+z_i^{-na})\Bigg)\prod_{1\leq i< j\leq N_f}|z_i-z_j|^2\nonumber 
\end{eqnarray}
\begin{eqnarray}
=(-1)^{-|\lambda|}S_{\lambda}(1^{N_f})\sum_{a=0}^{\infty}L_{n,a}\Big(\frac{[na]_{\lambda}}{[na+N_f]_{\lambda}}M_{N_f}(na,-na,1)+\frac{[-na]_{\lambda}}{[-na+N_f]_{\lambda}}M_{N_f}(-na,na,1)\Big).\nonumber\\
\end{eqnarray}
For the full correlation function, up to a numerical factor, we have 
\begin{eqnarray}
\C_{n}&=&\bra{\Uparrow}\prod_{i=1}^{N_f}\sigma^{+}_{j_i}e^{-t{\hat{H}}_{n}}\prod_{i=1}^{N_f}\sigma^{-}_{l_i}\ket{\Uparrow}\nonumber\\
&=&\frac{1}{(2\pi)^{N_f}}\prod_{i=1}^{N_f}\int_{-1}^{1}dz_iS_{\mu}(z^{-1})S_{\lambda}(z)\Bigg(\sum_{a=0}^{\infty}L_{n,a}(z_i^{na}+z_i^{-na})\Bigg)\prod_{1\leq i< j\leq N_f}|z_i-z_j|^2\nonumber
\end{eqnarray}
\begin{eqnarray}
=(-1)^{|\mu|-|\lambda|}S_{\mu}(1^{N_f})S_{\lambda}(1^{N_f})\sum_{a=0}^{\infty}L_{n,a}\Big(K_{N_f}(na; \mu, \lambda)+K_{N_f}(-na; \mu, \lambda)\Big).\nonumber\\
\end{eqnarray}

Next, we obtain the partition function of the generalized XX0 spin chain from the above expression,
\begin{eqnarray}
\label{General Z}
\Z_{Gen}&=&\bra{\Uparrow}\prod_{j=1}^{N_f}\sigma^{+}_{N_f-j}e^{-t{\hat{H}}_{Gen}}\prod_{i=1}^{N_f}\sigma^{-}_{N_f-i}\ket{\Uparrow}\nonumber\\
&=&\frac{1}{(2\pi)^{N_f}}\prod_{i=1}^{N_f}\int_{-1}^{1}dz_i\prod_{n=0}^{(N-1)/2}\Bigg(\sum_{a=0}^{\infty}L_{n,a}(z_i^{na}+z_i^{-na})\Bigg) \prod_{1\leq i< j\leq N_f}|z_i-z_j|^2\nonumber\\ 
&=&\frac{1}{(2\pi)^{N_f}}\sum_{a=0}^{\infty} \prod_{n=0}^{(N-1)/2}L_{n,a}\prod_{i=1}^{N_f}\int_{-1}^{1}dz_i\ z_i^{\frac{a(1-N^2)}{8}}(-1;z_i^{2a})_{\frac{N+1}{2}}\prod_{1\leq i< j\leq N_f}|z_i-z_j|^2\nonumber\\ 
&=&\sum_{a=0}^{\infty}L_{n,a}M_{N_f} \Big(\frac{a(1-N^2)}{8},-\frac{a(1-N^2)}{8},1\Big)\nonumber\\
&\quad\times&{}_2F_1\Big(-N_f,-\frac{a(1-N^2)}{8}; -N_f-\frac{a(1-N^2)}{8}; \big\{z_i^{2a(j-1)-1}\big\}_{j=1}^{(N+1)/2}\Big),\nonumber\\
\end{eqnarray}
where the q-Pochhammer symbol and hypergeometric function are defined by $(-1;z_i^{2a})_{\frac{N+1}{2}}\coloneqq\prod_{j=0}^{\frac{N-1}{2}}(1+z_i^{2a j})$ and ${}_rF_s(a_1, ..., a_r; b_1, ..., b_s; x_1, ..., x_m)\coloneqq \sum_{\lambda}\frac{[a_1]_\lambda...[a_r]_\lambda}{[b_1]_\lambda...[b_s]_\lambda}\frac{S_\lambda(x_1, ..., x_m)}{c_\lambda}$ with $c_\lambda\coloneqq\prod_{s\in\lambda}(a(s)+l(s)+1)$, and $a(s)$ and $l(s)$ are arm and leg of the square $s$ in the partition $\lambda$. From the second to the third line, we first changed the order of the product and sum and then used the identity $\prod_{n=0}^{(N-1)/2} (z^{an}+z^{-an})=z_i^{\frac{a(1-N^2)}{8}}(-1;z_i^{2a})_{\frac{N+1}{2}}$ and from third to the fourth line we used the Proposition (13.1.3) in \unscite{forrester}.

The current techniques do not allow for an explicit evaluation of the matrix integral of the generalized model with two or even one Schur function and thus we leave the evaluation of the correlation function of the generalized model as an open problem.

\section{Fredholm determinants in generalized XX0 spin chain}
In this part, we explain briefly, without derivations, the integrable probability properties of the original and generalized XX0 models. We briefly adopt the results, connecting discrete/continuous TD and FD, in the context of the generalized XX0 model. Our goal is to demonstrate the FD representation of the generalized XX0 model and the integrable probability that it brings into the picture. The actual extended and full-fledged implications and applications of this method for spin chains remain for future studies.

By putting together the following three results, we obtain the FD representation and thus the integrability property for the partition function of the finite-size generalized XX0 model: i) (continuous/discrete) TD representation of the XX0 partition function \unscite{Sa-Za}, ii) FD formula for the ratio of the discrete and continuous TD \unscite{Baik} and iii) FD representation of the continuous TD \unscite{Bo-Ok}. 

FD can be considered as a complex function that generalizes the determinant of a matrix. Formally, the FD of the kernel $\mathcal{G}(i,j)$ ($i,j \in \mathbb{Z}$ and restricted to $[n,\infty]$) can be defined as
\begin{equation}
\label{FD def}
\det{(1-\mathcal{G})}_{[n,n+1, ...]} \coloneqq  \sum_{m=0}^{\infty}(-1)^m \sum_{n\leq z_1<...<z_m}^{\infty}\det{\mathcal{G}\big((z_i,z_j)\big)_{i,j=1}^{m}}.
\end{equation}
The continuous and discrete TD are defined as
\begin{eqnarray}
\label{Teoplitz}
D_{N_{f}}({f})\coloneqq \det\begin{bmatrix}\int_{|z|=1}z^{-j+l}f(z)\frac{dz}{2\pi iz}\end{bmatrix}_{j,l=0}^{N_{f}-1},\quad D_{N_{f}}^{d}(f,|d|)\coloneqq \det\begin{bmatrix}\frac{1}{|d|} \sum_{z \in d} z^{-j+l}f(z)\end{bmatrix}_{j,l=0}^{N_{f}-1},
\end{eqnarray}
where $d$ is a finite domain with size $|d|$ and $f(z)$ is the weight function. 

As we have observed in chapter 2, by using Heine-Szeg\"o identity, the partition function of the generalized XX0 model has the TD representation,
\begin{eqnarray}
\label{Teoplitz Z}
\Z_{N_f}(f)=D_{N_{f}}(f),\quad  \Z_{N_f}^{d}(f, |d|)=D_{N_{f}}^{d}(f, |d|).
\end{eqnarray}
In the following, we present another determinant representation for the partition function, namely the FD, which is closely related to integrable/universal properties of the model and also provides a proper tool for studying the asymptotic limit of the models. In the next step, by using Borodin-Okounkov results \unscite{Bo-Ok}, that connects the TD to FD, we give the FD representation for the partition function of the generalized infinite model, 
\begin{equation}
\label{Fredholm for continuous}
\Z_{N_f}(f)=D_{N_{f}}(f)=\det(I-\mathcal{G}),
\end{equation}
where $f=\exp{(tV_{General})}$, and the integral operator $\mathcal{G}$ has the following kernel,
\begin{eqnarray}
\mathcal{G}(k,l)=(\frac{1}{2\pi \imath})^2\oint_{|w|=\rho<1}\oint_{|z|=\rho^{-1}>1}\frac{dz dw}{z^{k+1}w^{-l}}\frac{\exp{\big(V(z)-V(w)\big)}}{z-w}=\nonumber
\end{eqnarray}
\begin{eqnarray}
\label{kernel contin}
\frac{1}{k-l}(\frac{1}{2\pi \imath})^2\oint_{|w|=\rho<1}\oint_{|z|=\rho^{-1}>1}\frac{dz dw}{z^{k+1}w^{-l}}\frac{z(d/dz)V(z)-w(d/dw)V(w)}{z-w}\exp{\big(V(z)-V(w)\big)},\quad k\neq l\nonumber\\
\end{eqnarray}
where $V(z)=t\sum_{m=1}^{\frac{N-1}{2}}\Delta_m\Big(z^{-m}-z^{m}\Big)$.
This is an explicit determinantal formula which lies behind the exact solvability and universality of the model.

A natural question here is whether we have similar FD representation for the finite-size generalized XX0 spin chain. The answer is positive and in order to find the FD expression, we combine a recent result \unscite{Baik} and the product formula for FD determinants. Adopting, to our spin chain context, the FD formula appears in the ratio of discrete and continuous partition functions,
\begin{equation}
\label{Fredholm for ratio}
\frac{\Z_{N_f}^{d}(f, |d|)}{\Z_{N_f}(f)}=\frac{D_{N_{f}}^{d}(f,|d|)}{D_{N_{f}}(f)}=\det(I-\mathcal{K}),
\end{equation}
where the kernel of the integral operator $\mathcal{K}$ is given by
\begin{equation}
\label{F1}
\mathcal{K}(z,w)=z^{-N_f}\frac{p_{N_f}(z)p^{*}_{N_f}(w)-p^{*}_{N_f}(z)p_{N_f}(z)}{1-z^{-1}(w)}\sqrt{v_N(z)v_N(w)e^{V(z)}e^{V(w)}},
\end{equation}
and $p_{N_f}(z)$ and $p_{N_f}^*(z)=z^{N_f}\overline{p_{N_f}(\bar{z}^{-1})}$ are orthogonal polynomials with respect to the weight function $f=\exp{(t V_{General})}$, and the discrete function $v_N(z)$ in our domain ($z^N=1$) is defined by 
\begin{equation}
\label{F3}
v_N(z):=
  \begin{cases}
   -\frac{z^N}{1-z^N} & \quad z\in S^1_{in} \\
   \frac{z^{-N}}{1-z^{-N}} & \quad z\in S^1_{out} \\
  \end{cases},
\end{equation}
with contours $S^1_{in}$ and $S^1_{out}$ are positively-oriented circles of radii $1-\epsilon$ and
$1+\epsilon$ (for some infinitesimal positive $\epsilon$), respectively.

Using the ratio result in Eq.~\ref{Fredholm for ratio}, the continuous result Eq.~\ref{Fredholm for continuous}, and the product formula of the FD, we obtain the FD formula for the finite-size generalized XX0 model,
\begin{eqnarray}
\label{Fredholm for disc}
\Z_{N_{f}}^{d}(f,|d|)&=&\det(I-\mathcal{G})\det(I-\mathcal{K})\nonumber\\
&=&\det((I-\mathcal{G})(I-\mathcal{K}))=\det(I-\mathcal{H}),\quad \mathcal{H}=\mathcal{G}+\mathcal{K}-\mathcal{G}\mathcal{K}.
\end{eqnarray}
In order to apply the above abstract result \ref{Fredholm for disc}, we employ the above techniques in a concrete example and show how this leads to explicit results in spin chains. As an example we consider the case of the original XX0 model.
In this case, for the continuous partition function, we have \unscite{Joh},
\begin{eqnarray}
\label{Bessel partition}
\Z_{N_f}(f_{XX0})=D_{N_{f}}(f_{XX0})=e^{-t^2}\det(J_{k-l}(2\imath t))_{1\leq k,l\leq N_f}=\det(I-\mathcal{G}_{Be}),
\end{eqnarray}
where $J_i(z)$ is the Bessel function and the Bessel kernel is given by
\begin{eqnarray}
\label{Bessel kernel}
\mathcal{G}_{Be}(k,l;t)&=&\frac{t\big(J_k(2t)J_{l+1}(2t)-J_{k+1}(2t)J_{l}(2t)\big)}{k-l} \quad (k\neq l)\nonumber\\
&=&\sum_{n=1}^{\infty}J_{k+n}(2t)J_{l+n}(2t).
\end{eqnarray}
For the ratio of the discrete to the continuous partition function, as it is obtained in Gross-Witten matrix model \unscite{Baik},
\begin{eqnarray}
\label{Airy partition}
\frac{\Z_{N_f}^{d}(f_{XX0}, |d|)}{\Z_{N_f}(f_{XX0})}=\det(I-\mathcal{K}_{Ai}),
\end{eqnarray}
where $f_{XX0}=\exp{(t(z+z^{-1}))}$, and $\mathcal{K}_{Ai}$ is the Airy kernel defined by
\begin{eqnarray}
\label{Airy kernel}
\mathcal{K}_{Ai}(k,l)=\frac{Ai(k)Ai'(l)-Ai'(k)Ai(l)}{k-l},\quad Ai(x)=\frac{1}{2\pi}\int_{-\infty}^{\infty}e^{\imath s^3/3+\imath x s}ds.
\end{eqnarray}
Finally, for the discrete partition function of XX0 model using Eqs. \ref{Fredholm for disc}, \ref{Bessel partition} and \ref{Airy partition}, we obtain,
\begin{eqnarray}
\Z_{N_f}^{d}(f_{XX0}, |d|)=\det(1-\mathcal{H}_{Ai-Be}),\quad \mathcal{H}_{Ai-Be}= \mathcal{G}_{Be}+\mathcal{K}_{Ai}-\mathcal{G}_{Be}\mathcal{K}_{Ai}.
\end{eqnarray}

Having obtained the partition function of the infinite generalized model from the Selberg integral Eq. \ref{General Z}, then using the FD formula \ref{Fredholm for disc}, we can in principle obtain the partition function of the finite generalized model in terms of $M_{N_f}$. However,  further explicit studies of FD representation of the generalized XX0 model will be postponed to future studies.
Needless to say, the FD formulas for the partition functions of the finite/infinite generalized XX0 spin chain are also valid for all special cases of the generalized model, considering the appropriate potential of each case. Besides the exact closed expression for the partition function that the TD and FD provide, these results are also used in the actual computations in the asymptotic regime and extracting the phase structure \unscite{Baik}. In some special cases of the general potential, the limit of the FD tends to TW distribution \unscite{Baik2}, \unscite{Baik}. We continue our previous study \unscite{Sa-Za} and in this study, we use these methods and results to extract the phase structure of the generalized XX0 model.

\mybibitem{Sa-Za}{Saeedian, Meghdad, and Ali Zahabi. "Phase structure of XX0 spin chain and nonintersecting Brownian motion." Journal of Statistical Mechanics: Theory and Experiment 2018, no. 1 (2018): 013104. preprint arXiv:1612.03463}
\mybibitem{Schollwock}{Mikeska, Hans-Jürgen, Alexei K. Kolezhuk, Ulrich Schollwock, Johannes Richter, Damian JJ Farnell, and Raymod F. Bishop. "Quantum Magnetism." (2004).}
\mybibitem{Bogoliubov}{
Bogoliubov, N. M. "XX0 Heisenberg chain and random walks." Journal of Mathematical Sciences 138, no. 3 (2006): 5636-5643.}
\mybibitem{Its1}{Its, A. R., A. G. Izergin, and V. E. Korepin. "Temperature correlators of the impenetrable Bose gas as an integrable system." Communications in mathematical physics 129, no. 1 (1990): 205-222.}
\mybibitem{Its2}{Its, A. R., A. G. Izergin, V. E. Korepin, and N. A. Slavnov. "Differential equations for quantum correlation functions." International Journal of Modern Physics B 4, no. 05 (1990): 1003-1037.}
\mybibitem{Its3}{Its, A. R., A. G. Izergin, and V. E. Korepin. "Space correlations in the one-dimensional impenetrable Bose gas at finite temperature." Physica D: Nonlinear Phenomena 53, no. 1 (1991): 187-213.}
\mybibitem{Its4}{Its, A. R., A. G. Izergin, V. E. Korepin, and G. G. Varzugin. "Large time and distance asymptotics of field correlation function of impenetrable bosons at finite temperature." Physica D: Nonlinear Phenomena 54, no. 4 (1992): 351-395.}
\mybibitem{Izergin}{Izergin, A. G., A. R. Its, V. E. Korepin, and N. A. Slavnov. "Integrable differential equations for temperature correlation functions of the XXO Heisenberg chain." Journal of Mathematical Sciences 80, no. 3 (1996): 1747-1759.}
\mybibitem{Tracy2}{Tracy, Craig A., and Harold Widom. "Level-spacing distributions and the Airy kernel." Communications in Mathematical Physics 159, no. 1 (1994): 151-174.}

\mybibitem{Tracy}{Tracy, Craig A., and Harold Widom. "On orthogonal and symplectic matrix ensembles." Communications in Mathematical Physics 177, no. 3 (1996): 727-754.}
\mybibitem{Tracy3}{Tracy, Craig A., and Harold Widom. "Nonintersecting brownian excursions." The Annals of Applied Probability 17, no. 3 (2007): 953-979.}
\mybibitem{Bernard}{Bernard, Denis, Vincent Pasquier, and Didina Serban. "Exact solution of long-range interacting spin chains with boundaries." EPL (Europhysics Letters) 30, no. 5 (1995): 301.}
\mybibitem{Lipkin1}{Lipkin, Harry J., N. Meshkov, and A. J. Glick. "Validity of many-body approximation methods for a solvable model:(I). Exact solutions and perturbation theory." Nuclear Physics 62, no. 2 (1965): 188-198.}
\mybibitem{Lipkin2}{Meshkov, N., A. J. Glick, and H. J. Lipkin. "Validity of many-body approximation methods for a solvable model:(II). Linearization procedures." Nuclear Physics 62, no. 2 (1965): 199-210.}
\mybibitem{Lipkin3}{Glick, A. J., H. J. Lipkin, and N. Meshkov. "Validity of many-body approximation methods for a solvable model:(III). Diagram summations." Nuclear Physics 62, no. 2 (1965): 211-224.}
\mybibitem{Baik}{Baik, Jinho, and Zhipeng Liu. "Discrete Toeplitz/Hankel determinants and the width of nonintersecting processes." International Mathematics Research Notices (2013): rnt143.}
\mybibitem{Borodin}{Borodin, Alexei. "Determinantal point processes." arXiv preprint arXiv:0911.1153 (2009).}
\mybibitem{Forrester1}{Forrester, Peter, and S. V. E. N. Warnaar. "The importance of the Selberg integral." Bulletin of the American Mathematical Society 45, no. 4 (2008): 489-534.}
\mybibitem{Bo-Ok}{Borodin, Alexei, and Andrei Okounkov. "A Fredholm determinant formula for Toeplitz determinants." Integral Equations and Operator Theory 37, no. 4 (2000): 386-396.}

\mybibitem{Gross}{Gross, David J., and Edward Witten. "Possible third-order phase transition in the large-N lattice gauge theory." Physical Review D 21, no. 2 (1980): 446.}
\mybibitem{Gold}{Goldenfeld, Nigel. "Lectures on phase transitions and the renormalization group." (1992).}
\mybibitem{Zah}{Zahabi, Ali. "work in progress"}
\mybibitem{Tierz}{Pérez-García, David, and Miguel Tierz. "Chern–Simons theory encoded on a spin chain." Journal of Statistical Mechanics: Theory and Experiment 2016, no. 1 (2016): 013103.}
\mybibitem{Lieb} {E. Lieb, T. Schultz, D. Mattis, Ann. Phys. (NY) 16 (1961) 407.}
\mybibitem{Mehta org}{Mehta, M. L. Random Matrices (Pure and applied mathematics, v. 142). Edited by Madan Lal Mehta. Elsevier Science Limited, 2004.}
\mybibitem{Macdonald}{Macdonald, Ian Grant. Symmetric functions and Hall polynomials. Oxford university press, 1998.}
\mybibitem{Kad}{Kadell, Kevin WJ. "An integral for the product of two Selberg-Jack symmetric polynomials." Compositio Math 87 (1993): 5-43.}
\mybibitem{forrester}{Forrester, Peter J. Log-gases and random matrices (LMS-34). Princeton University Press, 2010.}
\mybibitem{Joh}{Johansson, Kurt. "Discrete orthogonal polynomial ensembles and the Plancherel measure." Annals of Mathematics 153, no. 1 (2001): 259-296.}

\mybibitem{Baik2}{Baik, Jinho, Percy Deift, and Kurt Johansson. "On the distribution of the length of the longest increasing subsequence of random permutations." Journal of the American Mathematical Society 12, no. 4 (1999): 1119-1178.}
\mybibitem{Ba-Di}{Baik, Jinho, Robert Buckingham, and Jeffery DiFranco. "Asymptotics of Tracy-Widom distributions and the total integral of a Painlevé II function." Communications in Mathematical Physics 280, no. 2 (2008): 463-497.}
\mybibitem{Mehta}{Mehta, Madan Lal. Random matrices. Vol. 142. Academic press, 2004.}
\mybibitem{Za}{Zahabi, Ali. "New phase transitions in Chern-Simons matter theory." Nuclear Physics B 903 (2016): 78-103.}
\mybibitem{Sc-Ma}{Majumdar, Satya N., and Gr\'egory Schehr. "Top eigenvalue of a random matrix: large deviations and third order phase transition." Journal of Statistical Mechanics: Theory and Experiment 2014, no. 1 (2014): P01012.}
\mybibitem{Dei}{Deift, Percy. "Universality for mathematical and physical systems." arXiv preprint math-ph/0603038 (2006).}
\mybibitem{Ju-Za}{Jurkiewicz, J., and Kasper Zalewski. "Phase structure of U ($N\rightarrow\infty$) gauge theory on a two-dimensional lattice for a broad class of variant actions." Nuclear Physics B 220, no. 2 (1983): 167-184.}
\mybibitem{Hinrichsen}{Hinrichsen, H. Non-equilibrium critical phenomena and phase transitions into absorbing states. Advances in physics 49, 815–958 (2000)}

\mybibitem{Fo-Ma}{Forrester, Peter J., Satya N. Majumdar, and Gr\'egory Schehr. "Non-intersecting Brownian walkers and Yang–Mills theory on the sphere." Nuclear Physics B 844, no. 3 (2011): 500-526.}

\mybibitem{Do-Ka}{Douglas, Michael R., and Vladimir A. Kazakov. "Large N phase transition in continuum QCD2." Physics letters B 319, no. 1-3 (1993): 219-230.}

\mybibitem{Pe-Sh 1}{Periwal, Vipul, and Danny Shevitz. "Unitary-matrix models as exactly solvable string theories." Physical review letters 64, no. 12 (1990): 1326.}

\mybibitem{Pe-Sh 2}{Periwal, Vipul, and Danny Shevitz. "Exactly solvable unitary matrix models: multicritical potentials and correlations." Nuclear Physics B 344, no. 3 (1990): 731-746.}

\mybibitem{Le-Ma}{Le Doussal, Pierre, Satya N. Majumdar, and Gr\'egory Schehr. "Multicritical edge statistics for the momenta of fermions in nonharmonic traps." Physical review letters 121, no. 3 (2018): 030603.}

\mybibitem{Ste}{St\'ephan, Jean-Marie. "Free fermions at the edge of interacting systems." arXiv preprint arXiv:1901.02770 (2019).}

\mybibitem{Ab-Ig}{Abarenkova, Nina Igorevna, and Andrei Georgievich Pronko. "Temperature correlation function in the absolutely anisotropic XXZ Heisenberg magnet." Theoretical and mathematical physics 131, no. 2 (2002): 690-703.}

\mybibitem{Bo-Ma 2}{Bogoliubov, N. M., and C. L. Malyshev. "Ising limit of a Heisenberg XXZ magnet and some temperature correlation functions." Theoretical and Mathematical Physics 169, no. 2 (2011): 1517-1529.}

\mybibitem{Bo-Iz}{Bogoliubov, N. M., A. G. Izergin, and N. A. Kitanine. "Correlation functions for a strongly correlated boson system." Nuclear Physics B 516, no. 3 (1998): 501-528.}

\mybibitem{Bo-Ma rev}{Bogolyubov, Nikolai Mikhailovich, and Cyril Leonidovich Malyshev. "Integrable models and combinatorics." Russian Mathematical Surveys 70, no. 5 (2015): 789.
}

\mybibitem{Bo-Ma com}{Bogoliubov, N., and Cyril Malyshev. "Correlation Functions as Nests of Self-Avoiding Paths." Journal of Mathematical Sciences 238, no. 6 (2019): 779-792.}

\mybibitem{Cl-Kr}{Claeys, Tom, Igor Krasovsky, and Alexander Its. "Higher‐order analogues of the Tracy‐Widom distribution and the Painlevé II hierarchy." Communications on pure and applied mathematics 63, no. 3 (2010): 362-412.}

\mybibitem{Ak-At}{Akemann, Gernot, and Max R. Atkin. "Higher order analogues of Tracy–Widom distributions via the Lax method." Journal of Physics A: Mathematical and Theoretical 46, no. 1 (2012): 015202.}

\end{document}